\documentclass[11pt]{article}
\usepackage{eurosym}
\usepackage{amsmath}
\usepackage{amsfonts}
\usepackage{amssymb}
\usepackage{graphicx}
\usepackage{ulem}
\usepackage{slashed}
\usepackage{color}

\begin{document}

\title{\textbf{Chern-Simons Forms in Gravitation Theories}$^*$}
\author{\textbf{Jorge Zanelli} \\
Centro de Estudios Cient\'{\i}ficos (CECs), Casilla 1469, Valdivia, Chile\\
Universidad Andr\'es Bello, Rep\'ublica 440, Santiago, Chile}
\date{}
\maketitle

\begin{abstract}
The Chern-Simons (\textbf{CS}) form evolved from an obstruction in mathematics into an important object in theoretical physics. In fact, the presence of CS terms in physics is more common than one may think: they seem to play an important role in high Tc superconductivity and in recently discovered topological insulators. In classical physics, the minimal coupling in electromagnetism and to the action for a mechanical system in Hamiltonian form are examples of CS functionals.  CS forms are also the natural generalization of  the minimal coupling between the electromagnetic field and a point charge when the source is not point-like but an extended fundamental object, a membrane. They are found in relation with anomalies in quantum field theories, and as Lagrangians for gauge fields, including gravity and supergravity.  A cursory review of the role of CS forms in gravitation theories is presented at an introductory level.

$^*$This is an author-created, un-copyedited version of an article accepted for publication in Classical and Quantum Gravity \textbf{29} 133001 (2012). IOP Publishing Ltd is not responsible for any errors or omissions in this version of the manuscript or any version derived from it. The definitive publisher-authenticated version is available online at  doi:10.1088/0264-9381/29/13/133001.
\end{abstract}

\section{Introduction}  

The \textit{discovery} of CS forms in mathematics was like a geographical discovery: A frustrated effort to find a particular formula led to the discovery of an unexpected obstruction, a ``boundary term", an object that could be locally written as a total derivative, but not globally. In their foundational paper, Shiin-Shen Chern and James Simons describe their discovery as follows \cite{Chern+Simons}:\\
\textit{This work} [...] \textit{grew out of an attempt to derive a purely combinatorial formula for the first Pontrjagin number} [...]. \textit{This process got stuck by the emergence of a boundary term which did not yield to a simple combinatorial analysis. The boundary term seemed interesting in its own right and it and its generalization are the subject of this paper.}

Four decades after \cite{Chern+Simons}, CS forms have opened new areas of study in mathematics and several excellent books aimed at their many applications in physics have been written \cite{Nakahara,Baez}. These notes are no substitute for them. Our purpose here is to merely collect a few useful observations that could help understand the role of CS forms in physics and the reason for their usefulness. In this spirit, these notes should not be taken as an exhaustive discussion of all the relevant aspects of Chern-Simons forms in gravitation. This is an elementary introduction, discussing a selection of topics related to gravity, that is expected to be self-contained and pedagogical. There is an abundant literature describing solutions and other applications of CS gravities, like \cite{BHscan,Charmousis,Garraffo}. Here, the emphasis is on the construction of the action principles, and in the geometric features that make CS forms particularly suited for the physics of geometric systems.

These notes are based on several lectures given at schools over the past decade, and they will hopefully become part of a forthcoming book on the same subject.

\textbf{Road map}

This first section is devoted to describe CS forms and why they are useful --and widely used-- in physics both as actions for gauge theories and as couplings to sources, including the standard interaction between the electromagnetic field and a point charge. In (1.3), the gauge invariance of gravity is reviewed, emphasizing the distinction between coordinate transformations --trivial relabelling of clocks and measuring sticks-- and the true gauge symmetry: local Lorentz invariance.

Section 2 describes the basic ingredients of gravity in the first order formalism, while section 3 takes those fields as building blocks to produce a huge family of gauge invariant gravitational actions beyond the modest Einstein-Hilbert form. This over-abundance of theories is trimmed in section 4 by imposing a disciplinary condition: the arbitrary coefficients in the Lagrangian are chosen in such a way that the gauge symmetry becomes enlarged. This leads in odd dimensions to an almost unique choice of gauge theories with $SO(D,1)$ (de Sitter), $SO(D-1,2)$ (anti-de Sitter), or $ISO(D-1,1)$ (Poincar\'e),  symmetry groups. 

Section 5 deals with the role of CS forms as couplings to sources, as they naturally serve to introduce interactions between the spacetime geometry and topological structures such as angular defects (naked point like singularities, or zero-branes), or extended localized sources such as $2p$-branes.  

Section 6 summarizes how the construction is naturally extended to supergravity theories by enlarging the gauge group to a supergroup. The fields now become components of a connection in a superalgebra, thus producing locally supersymmetric theories coupled to gravity as gauge theories. Again, the structure that makes possible this action is the CS form for the corresponding (super) connection, a construction that is only possible in odd dimensions. 

\subsection{Why Chern-Simons forms?}   

The key feature of CS forms that makes them useful in physics stems from the fact that they are quasi-invariant under gauge transformations: If $\mathbf{A}$ is the Yang-Mills vector potential (connection) for a nonabelian gauge field theory, under a gauge transformation 
\begin{equation}
\mathbf{A}_{\mu}(x) \rightarrow \mathbf{A}'_{\mu}(x) = g^{-1}(x)[\mathbf{A}_{\mu}(x) +d]g(x),
\label{connection}
\end{equation}
where $g(x)$ is a group element that can be continuously deformed to the identity everywhere, then a CS form $\mathcal{C}(\mathbf{A})$ transforms as
\begin{equation}
\mathcal{C}(\mathbf{A}')= \mathcal{C}(\mathbf{A}) + d\Omega.
\end{equation}
In other words, even if the group is nonabelian, the CS form transforms as an abelian connection. 

There are two instances in classical physics where a function that changes by a total derivative --like an abelian gauge field--, appear to have nontrivial consequences:

$\bullet$ In the coupling between the electromagnetic vector potential $A_\mu$ and a current made up of point charges, and 

$\bullet$ In the transformation of a Lagrangian under a symmetry transformation.

In fact, these two are not completely independent situations. The existence of a symmetry gives rise through Noether's theorem to a conserved current; this current in turn couples to the dynamical variables as a source for the classical equations. This interplay can be verified in the case of the gauge invariance of Maxwell's equations and the conserved Noether current made up of point electric charges. 

The lesson one draws from this is that if a $p$-form $\mathcal{C}$ changes by a total derivative under a group of transformations ($\mathbb{G}$) then $\mathcal{C}$ is a good candidate for a gauge invariant Lagrangian, and for something that couples to a gauge invariant source defined by a conserved current. Indeed, if one couples $\mathcal{C}$ to a $\mathbb{G}$-invariant current $j$, through the form 
\begin{equation}
I=\int \mathcal{C}\wedge *j,
\label{minimal-c}
\end{equation}
then, under $\mathbb{G}$, $I$ changes as
\begin{equation}
\delta I=\int_M \delta (\mathcal{C})\wedge *j=\int_M (d\Omega)\wedge *j =\int_M d(\Omega \wedge *j)+(-1)^p \int_M \Omega \wedge d*j.
\label{delta-I}
\end{equation}
where $\Omega$ is an arbitrary ($p-1$)-form. The first term in the right hand side can be turned into a a surface integral on $\partial M$ that can be dropped for sufficiently localized $\Omega$ and/or $j$. The critical term is the last one, which vanishes for any generic $\Omega$ if and only if $d*j=0$. Hence, the ``minimal coupling" (\ref{minimal-c}) is $\mathbb{G}$-invariant provided the source to which $\mathcal{C}$ couples is conserved.

The reader might have guessed by now that $\mathcal{C}$ will be a CS form, and $d*j=0$ will be the conservation law associated to the sources. 

\subsection{Constructing CS forms}  

A CS form is defined by two ingredients: a symmetry group $\mathbb{G}$ in a  certain representation, and an odd-dimensional manifold $M$ over which one can define functions and fields, that is, a differentiable manifold. The fundamental object in a gauge theory is the gauge \textit{connection}, a generalization of the abelian vector potential, $A=A_{\mu}dx^{\mu}$, a matrix-valued field one-form,\footnote{In what follows, differential forms will be used throughout unless otherwise indicated. We follow the notation and conventions of \cite{Schutz}.}
\begin{eqnarray}
\mathbf{A}&=&\mathbf{A}_{\mu}dx^{\mu}\\ \nonumber
&=&A^a_{\mu}\mathbf{K}_a dx^{\mu},
\end{eqnarray}
where $\mathbf{K}_a$, $a=1, 2, ..., N$ are generators of the gauge group $\mathbb{G}$. The connection is defined to transform in the adjoint representation of the gauge group. If $g(x)=\exp[\alpha^a(x) \mathbf{K}_a]$ is an element of the group, it acts on the connection as
\begin{equation} 
\mathbf{A}\xrightarrow{g} \mathbf{A}'=g\mathbf{A}g^{-1}+gd(g^{-1}) .
\label{A'}
\end{equation}
This transformation law is defined by the need to set up a covariant derivative, $D=d-\mathbf{A}$, that generalizes $\partial_{\mu} -iA_{\mu}$ in electrodynamics, or $\nabla_{\mu}=\partial_\mu+\mathbf{\Gamma}_{\mu}$ in Riemannian geometry.

The connection is gauge-dependent and therefore not directly measurable. However, the curvature (field strength) $\mathbf{F}= d\mathbf{A} +\mathbf{A}\wedge\mathbf{A}$, transforms homogeneously,
\begin{equation}
\mathbf{F} \xrightarrow{g}  \mathbf{F}'=g\mathbf{F}g^{-1},
\label{F'}
\end{equation}
and, like  $\vec{E}$ and $\vec{B}$, has directly observable local features. For example, the $2k$-form 
\begin{equation}
Tr(\mathbf{F}^k),  
\end{equation}
is invariant under (\ref{F'}) by construction and therefore observable. Invariants of this kind (or more generally, the trace of any polynomial in $\mathbf{F}$), like the Euler or the  Pontryagin forms, are called \textit{characteristic classes}, which capture the topological nature of the mapping between the spacetime manifold and the gauge group, $\mathbf{A}: M \mapsto \mathbb{G}$.

Let's denote by $P_{2k}(\mathbf{F})= \langle \mathbf{F}^k \rangle$ one of those invariants, where $\langle \cdots \rangle$ stands for a symmetric, multilinear operation in the Lie algebra, a generalized trace in the algebra. In order to fix ideas, one can take it to be the ordinary trace, but some Lie algebras could have more than one way to define ``$Tr$" (the algebra of rotations, for example, has two). Then, $P_{2k}$ satisfies the following conditions \cite{Nakahara}:\\
\textbf{i}. It is a polynomial in the curvature $\mathbf{F}$ associated to a gauge connection $\mathbf{A}$.\\
\textbf{ii}. It is invariant under gauge transformations (\ref{A'}) and (\ref{F'}). \\
\textbf{iii}. It is closed, $dP_{2k}=0$. \\
\textbf{iv}. It can be locally expressed as the derivative of a $(2k-1)$-form, $P_{2k}=d\mathcal{C} _{2k-1}$.  \\
\textbf{v}. Its integral over a $2k$-dimensional compact, orientable manifold without boundary, is a topological invariant, $\int_M P_{2k}= c_{2k}(M) \in \textbf{Z}$.

Condition (\textbf{ii}) is satisfied by virtue of the cyclic property of the product of curvature 2-forms. Condition  (\textbf{iii}) is a consequence of the Bianchi identity which states that the covariant derivative (in the connection $\mathbf{A}$) of the curvature $\mathbf{F}$ vanishes identically, $D\mathbf{F}=d\mathbf{F}+[\mathbf{A},\mathbf{F}]\equiv 0$. Condition (\textbf{iv}) follows from (\textbf{iii}) as a direct consequence of Poincar\'e's lemma: \textit{If $d\phi=0$ then $\phi=d$(something)}. Finally, (\textbf{v}) means that, although $P_{2k}$ looks like an exact form in a local chart, globally it is not. 

The CS forms $\mathcal{C} _{2n-1}$ identified in \cite{Chern+Simons} are given by the trace of some polynomial in $\mathbf{A}$ and $d\mathbf{A}$ that \textit{cannot} be written as local functions involving only the curvature $\mathbf{F}$. This makes the CS forms rather cumbersome to write, but its exact expression in not needed in order to establish its most important property, as stated in the following

\textbf{Lemma}: \textit{Under a gauge transformation} (\ref{A'}), $\mathcal{C} _{2n-1}$ \textit{changes by a locally exact form (a total derivative in a coordinate patch).} 

\textbf{Proof:} The homogeneous polynomial $P_{2k}$ is invariant under gauge transformations (this is easily seen from the transformation (\ref{F'}) if $P_{2k}= Tr[\mathbf{F}^k]$, due to the cyclic property of the trace). Performing a gauge transformation on (\textbf{iv}),  gives
\begin{equation}
\delta_{gauge} P_{2k}=d(\delta_{gauge}\mathcal{C} _{2k-1}) , 
\end{equation}
and since $P_{2k}$ is invariant, one concludes that the right hand side must vanish as well,
\begin{equation}
d(\delta_{gauge}\mathcal{C} _{2k-1}=0) .
\end{equation}
By Poincar\'e's lemma, this last equation implies that the gauge variation of $\mathcal{C}_{2k-1}$ can be written locally as an exact form,
\begin{equation}
\delta_{gauge}\mathcal{C} _{2k-1}=d\Omega . \;\; \blacksquare
\label{CSform}
\end{equation}

This is a nontrivial result: although the nonabelian connection $\mathbf{A}$ transforms inhomogeneously, as in (\ref{A'}), the CS form transforms in the same way as an abelian connection. This is sufficient to ensure that a CS ($2n-1$)-form defines gauge invariant action in a ($2n-1$)-dimensional manifold,
\begin{equation}
\delta_{gauge} I[A]=\int_{M^{2n-1}} \delta_{gauge} \mathcal{C} _{2n-1} =\int_{M^{2n-1}} d\Omega ,
 \end{equation}
which vanishes for an appropriate set of boundary conditions.  

CS actions are exceptional in physics because, unlike for most theories, such as Maxwell or Yang-Mills, they do not require a metric structure. The gauge invariance of the action does not depend on the shape of the manifold $M^{2n-1}$; a metric structure may not even be defined on it. This is a welcome feature in a gravitation theory in which the geometry is dynamical. A particular consequence of this is that in CS gravity theories, the metric is a derived (composite) object and not a fundamental field to be quantized. This in turn implies that concepts such as the energy-momentum tensor and the inertial mass must be regarded as phenomenological constructs of classical or semi-classical nature, an emerging phenomenon.  

\subsection{The gauge invariance of gravity}  

In order to see how CS forms can enter in gravity, we must identify what is the gauge invariance associated to the gravitational interaction.

\subsubsection{Coordinate transformations}  

It is often stated that the fundamental symmetry of gravitation theory is the group of general coordinate transformations, the diffeomorphism group, as is usually called. These transformations, however, are not a useful symmetry and much less are they a unique feature of gravity. Indeed, any action, for any physical system whatsoever, is coordinate invariant. Therefore, all meaningful statements derived from an action principle must be coordinate invariant. In fact, \textit{all} well defined physical theories must be invariant under general changes of coordinates. 

It is a triviality that an objective situation cannot depend on the coordinates we humans employ to describe them. The \textit{representation} may of course change, but the phenomenon itself cannot. This is explicitly recognized in Lagrangian mechanics, where the choice of coordinates is left completely free. In other words, general coordinate transformations are not a distinctive symmetry of gravity, it is the invariance of the laws of Nature under the changes of human's forms of describing them.

Coordinates are introduced for the convenience of physicists when solving equations, but they are not a feature of the physical system. We experience this every time we write or Maxwell's equations or the Schr\"odinger operator in spherical coordinates, in order to render more transparent the presence of boundaries or sources with spherical symmetry. Here the coordinates are adapted to the symmetry of the physical situation, but that does not imply that coordinates could not be chosen otherwise. There is nothing sacred about coordinates beyond convenience. 

Diffeomorphisms can be conceived as the result of gauging the translation group, roughly described as local translations,
\begin{equation}
x^{\mu} \rightarrow x'^{\mu} =x^{\mu}  + \xi^{\mu}(x). 
\label{diffeos}
\end{equation}
These are gauge-like transformations in the sense that $\xi^{\mu}$ is an arbitrary function of $x$, but here the analogy with with gauge transformations stops. Under coordinate transformations, a vector transforms as 
\begin{equation}
v'^\mu(x')= L^\mu_\nu (x) v^\nu (x), \mbox{  where }  L^\mu_\nu (x)=\frac{\partial x'^{\mu} }{\partial x^{\nu}} = \delta^\mu_\nu+\frac{\partial \xi^{\mu} }{\partial x^{\nu}} .
\label{v'}
\end{equation}
In a gauge transformation the argument of the left hand side would be $x$ and not $x'$, which, according to the interpretation of (\ref{diffeos}) as a translation, corresponds to a different point. One could try to write (\ref{v'}) in a form similar to a gauge transformation,
\begin{eqnarray}\nonumber
v'^\mu(x)&=& L^\mu_\nu (x) v^\nu (x)-\xi^{\lambda}(x)\partial_\lambda v^\mu(x) \\ 
&=& v^\mu(x)+\frac{\partial \xi^{\mu} }{\partial x^{\nu}} v^\nu (x) -\xi^{\lambda}(x)\partial_\lambda v^\mu(x)\
\end{eqnarray}

The first two terms on the right of this last expression correspond to the way a vector representation transforms under the action of a gauge group in a fibre bundle. The last term, however, represents the drift, produced by the fact that the translation actually shifts the point in the manifold. This type of term is not present in a gauge transformation of the type discussed above, which means that the diffeomorphism group does not act as a local symmetry in a standard gauge theory, like Yang-Mills. 

A standard fibre bundle is locally a direct product of a base manifold and a group, each fibre being a copy of the group orbit. In the case of diffeomorphism group, the fibres lie along the base. Therefore this structure is not locally a product and the group of coordinate transformations on a manifold do not define a fibre bundle structure. Basically, the problem is that the translation group does not take a field at a given point into a different field at the same point, but changes the arguments of the fields, which is something gauge transformations never do, as can be seen in (\ref{connection}). 

Apart from the obvious fact that the group of translations is a rather trivial group, whose gauging could hardly describe the richness of gravity, it is apparent that the translation symmetry is violated by the curvature of spacetime itself. This wouldn't happen in a genuine gauge theory --gauge invariance is respected by all solutions of Maxwell's equations, and by all conceivable off-shell fields in quantum mechanics. This is a key feature that makes gauge symmetries extremely useful in the quantum description: the invariance is inherent to the field representation and is not spoiled by dynamics, be it classical or quantum.

The generators of diffeomorphisms $\mathcal{H}_\mu$ form an algebra whose Poisson brackets are 
\begin{equation}
\begin{array}{lll}
\lbrack {\cal H}_{\perp }(x),{\cal H}_{\perp }(y)] & = & g^{ij}(x) \delta(x,y),_{i}{\cal H}_{j}(y)-g^{ij}(y)\delta (y,x),_{i}{\cal H}_{j}(x) \\
\lbrack {\cal H}_{i}(x),{\cal H}_{j}(y)] & = & \delta(x,y),_{i}{\cal H}_{j}(y)-\delta (x,y),_{j}{\cal H}_{i}(y) \\
\lbrack {\cal H}_{\perp }(x),{\cal H}_{i}(y)] & = & \delta(x,y),_{i}{\cal H}_{\perp }(y).
\end{array},  \label{Hs}
\end{equation}

Note that this algebra is defined by a set of \textit{structure functions}  rather \textit{structure constants}, as in ordinary Lie algebras. This type of structure is not a Lie algebra, but an \textit{open algebra} \cite{Henneaux}, and what is a more serious concern, in gravity, the structure functions involve the metric of the manifold, which is itself a dynamical variable. This would represent a major drawback in a quantum theory, as the nature of the symmetry would be determined at every point of the manifold by the local geometry, which is itself a function of the observable operators of the theory. 

\subsubsection{Lorentz transformations} 

In 1907, Einstein observed that the effect of gravity can be neutralized by free fall. In a freely falling laboratory, the effect of gravity can be eliminated. This trick is a local one: the lab has to be small enough and the time span of the experiments must be short enough. Under these conditions, the experiments will be indistinguishable from those performed in absence of gravity, and the laws of physics that will be reflected by the experiments will be those valid in Minkowski space. This means that, in a local neighbourhood, spacetime possesses Lorentz invariance. In order to make manifest this invariance, it is necessary to perform an appropriate coordinate transformation to a particular reference system, viz., a freely falling one.
Conversely, Einstein argued that in the absence of gravity, the gravitational field could be mocked by applying an acceleration to the laboratory. This idea is known as the principle of equivalence meaning that, in a small spacetime region, gravitation and acceleration are equivalent effects.

A freely falling observer defines a local inertial system. For a small enough region around him or her, the trajectories of projectiles (freely falling as well) are straight lines and the discrepancies with Euclidean geometry are negligible. Particle collisions mediated by short range forces, such as those between billiard balls, molecules or subnuclear particles, satisfy the conservation laws of energy and momentum that are valid in special relativity.

The Equivalence Principle states that, in the presence of gravity, physical phenomena in a small vicinity (in space and time) of a freely falling observer cannot be distinguished from those in an inertial frame and in the absence of gravity. The fact that inertial observers moving at constant velocity with respect to each other would have the same physical laws, means that physical phenomena in a small neighbourhood of any spacetime point should be invariant under Lorentz transformations. Since these Lorentz transformations can act independently at each point indicates that the true gauge invariance of gravity is local Lorentz symmetry. Hence, Einstein's observation that the principle of equivalence is a central feature of general relativity, and this makes gravitation a gauge theory for the group $SO(3,1)$, the first nonabelian gauge theory ever proposed \cite{OR-S}.

Note that the Lorentz group can act independently at each point, unlike the translations which is a symmetry only in maximally symmetric spacetimes. The invariance of gravitation theory under $SO(3,1)$ is a minimal requirement, the complete group of invariance could be larger, $\mathbb{G}\supseteq SO(3,1)$. Natural options are the de Sitter ($SO(4,1)$), anti-de Sitter ($SO(3,2)$), conformal ($SO(4,2)$) and Poincar\'e $ISO(3,1)$) groups, or some of their supersymmetric extensions. The $SO(3,1)$ symmetry can also be seen as a remnant of a dimensional reduction from higher dimensions with a larger group like $SO(m,n)$, or even a supergroup from a fundamental higher-dimensional theory.

\section{Action principle} 

In order to implement this symmetry in an action principle, it is necessary to describe the spacetime geometry in terms of fields that correspond to some nontrivial representation of $SO(D-1,1)$, where from now on, $D$ denotes the spacetime dimension. This is most effectively done if the metric and affine features of the geometry are treated independently, which is obtained in an intuitively simple way introducing the notion of \textit{tangent space}. This approach is known as the first-order formalism because it produces first order field equations for gravity.

The first order formalism uses the exterior calculus of differential forms. It is sometimes held that the advantage of using differential forms is the compactness of the expressions as compared with the standard tensor calculus in coordinate bases \cite{Eguchi-Gilkey-Hanson}. Although this is certainly an advantage, it is also true that coordinates are still necessary in order to solve the Einstein's equations, so the advantage would be for the elegance in the formulation of the theory at best. 

The profound advantage of the first order formalism, however, is in the decoupling between the gauge symmetry of gravity (local Lorentz invariance) and the particular configurations of the spacetime geometry. It is the fact that the symmetry generators form a Lie algebra --with structure constants--, and not an open algebra like (\ref{Hs}), with structure functions that depend on the dynamical variables of the theory, that makes the first order formalism interesting and attractive.

\subsection{First order formalism} 

As discussed in the Appendix, the spacetime geometry can be captured by two fundamental fields, the vielbein $e^a_\mu(x)$ characterizing the metric structure, an the Lorentz connection, $\omega_{\;b\mu}^{a}(x)$, that codifies the affine features.\footnote{This section is a free interpretation of the ideas the author learned from lectures by B. Zumino \cite{Zumino}  and T. Regge \cite{Regge}, who in turn elaborated on earlier work by R. Utiyama \cite{Utiyama} and T. W. Kibble \cite{Kibble}.} Until further notice, these two fields are totally arbitrary and independent. The metric is a derived expression and not a fundamental field to be varied in the action given in (\ref{metric}). 

It can be observed that both the vielbein and the Lorentz connection appear as local 1-forms.
\begin{equation}
e^{a}\equiv e_{\mu}^{a}(x)dx^{\mu}  \quad \mbox{and} \quad  \omega _{\;b}^{a}\equiv \omega _{\;b\mu}^{a}(x)dx^{\mu}. \label{1Forms}
\end{equation}

Moreover, all geometric properties of $M$ can be obtained from these two 1-forms and their exterior derivatives only. Since both $e^{a}$ and $\omega^{a}_{\;b}$ carry only Lorentz indices but no coordinate indices ($\mu$,  $\nu$, etc.), these 1-forms, like all exterior forms, are invariant under coordinate transformations of $M$. This is why a description of the geometry that only uses these forms and their exterior derivatives, is naturally coordinate-free and trivially coordinate invariant.

In this formalism the spacetime tensors are replaced by tangent space tensors. In particular, the curvature is a Lorentz tensor,
\begin{eqnarray}
R_{\;b}^{a} &=& d\omega _{\;b}^{a}+\omega _{\;c}^{a}\wedge \omega_{\;b}^{c} \nonumber \\
&=&\frac{1}{2}R_{\;b\mu \nu }^{a}dx^{\mu}\wedge dx^{\nu}. \label{curvature}
\end{eqnarray}
This tensor 2-form measures how much a vector in the tangent space rotates when parallel-transported around an infinitesimal loop of area $dx^\mu \wedge dx^\nu$. In a space of vanishing torsion, this curvature two-form is related to the Riemann tensor, $R^{ab}_{\;\;\;\mu \nu}=e_{\;\alpha }^{a}e^{b}_{\;\;\beta}R^{\alpha\beta}_{\;\;\;\; \mu \nu}$.

\subsection{Building blocks}   

The fact that $\omega_{\;b}^{a}$ and the gauge potential $A_{\;b}^{a}=A_{\;b\mu }^{a}dx^{\mu }$ in Yang-Mills theory, are both 1-forms and have similar properties was noted long ago \cite{Utiyama}. They are both connections of a gauge group,  their transformation laws have the same form and the curvature (\ref{curvature}) and $F_{\;b}^{a}=dA_{\;b}^{a}+A_{\;c}^{a}\wedge A_{\; b}^{c}$ are completely analogous.\footnote{In more formal terms, $\omega$ and $A$ are locally defined Lie algebra valued 1-forms on $M$, $\omega$ in the principal $SO(D-1,1)$-bundle, and $A$ in the vector bundle of the group $G$, respectively.}

There is an asymmetry with respect to the vielbein, though. Its transformation properties under the Lorentz group is not that of a connection but of a vector, and the corresponding field in an arbitrary gauge theory would be a matter field. Another important geometric object obtained from derivatives of $e^{a}$ is the Torsion 2-form, $T^a=de^a+\omega^a_{\;b}\wedge e^b$, which is a function of both $e$ and $\omega$, while $R_{\;b}^{a}$ is not a covariant derivative of anything and a function of $\omega$ only.

Thus, the basic building blocks of first order gravity are $e^a$, $\omega^a_{\;b}$, $R^a_{\;b}$, $T^a$. There are no more building blocks, and with them we must put together an action. We are interested in objects that transform in a controlled way under Lorentz rotations (vectors, tensors, spinors, etc.). The existence of Bianchi identities implies that differentiating these fields, the only tensors that can be produced are combinations of the same objects. In the next sections we discuss the construction of the possible actions for gravity using these ingredients. 

\section{Gravity actions}   

We now turn to the construction of a gravity action. We expect it to be a local functional of the one-forms $e^{a}$, $\omega _{\;b}^{a}$ and their exterior derivatives. In addition, the two invariant tensors of the Lorentz group, $\eta _{ab}$, and $\epsilon _{a_{1}\cdot \cdot \cdot \cdot a_{D}}$ can be used to raise, lower and contract indices. We need not worry about invariance under general coordinate transformations as exterior forms are coordinate invariant by construction.

The use of only exterior products of forms excludes the metric, its inverse and the Hodge $\star $-dual (see \cite{Zumino} and \cite{Regge} for more on this). This postulate also excludes tensors like the Ricci tensor\footnote{Here $E_{a}^{\lambda}$ is the inverse vielbein, $E_{a}^{\lambda}e^b_{\; \lambda}=\delta ^a_b$ .} $R_{\mu \nu}=E_{a}^{\lambda}\eta_{bc}e_{\mu}^{c}R_{\; \lambda \nu}^{ab}$, or $R_{\alpha \beta}R_{\mu \nu}R^{\alpha \mu \beta \nu}$, except in very special combinations like the Gauss-Bonnet form, that can be expressed as exterior products of forms.

The action principle cannot depend on the choice of basis in the tangent space and hence Lorentz invariance should be ensured. A sufficient condition to have Lorentz invariant field equations is to demand the Lagrangian itself to be Lorentz invariant, but this is not really necessary. Allowing for the Lagrangian to be \textit{quasi-invariant} so that it changes by a total derivative --and the action changes by a boundary term--, still gives rise to covariant field equations in the bulk.

\subsection{Lorentz invariant Lagrangians}    
Let us consider first Lorentz invariant  Lagrangians. By inspection, one concludes that it must be a $D$-form consisting of linear combinations of products of $ e^{a},\;  R_{\; b}^{a}, \;T^{a}$, contracted with $\eta_{ab}$ and $\epsilon _{a_1\cdots a_D} $, and no $\omega$ \cite{Mardones-z}, 
\begin{eqnarray}\label{P}
\mathcal{P}_{2k}\!&=:& R^{a_1}{}_{a_2}  R^{a_2}{}_{a_3} \cdots R^{a_k}{}_{a_1}  \\\label{u}
\upsilon_k \!&=:& e_{a_1}R^{a_1}{}_{a_2}  R^{a_2}{}_{a_3} \cdots R^{a_k}{}_b e^b, \;\; \mbox{odd } k \\\label{t}
\tau_k\!&=:& T_{a_1}R^{a_1}{}_{a_2}  R^{a_2}{}_{a_3} \cdots R^{a_k}{}_b T^b, \;\; \mbox{even } k \\\label{z}
\zeta_k \!&=:&  e_{a_1}R^{a_1}{}_{a_2}  R^{a_2}{}_{a_3} \cdots R^{a_k}{}_b T^b \\\label{E}
\mathcal{E}_{D}\!&=:& \epsilon_{a_1 a_2\cdots a_D} R^{a_1 a_2} R^{a_3 a_4} \cdots R^{a_{D-1} a_D} , \;\; \mbox{even } D\\ \label{L}
L_p\!&=:& \epsilon_{a_1 a_2 \cdots a_{D}} R^{a_1 a_2} R^{a_3 a_4} \cdots R^{a_{2p-1} a_{2p}} e^{a_{2p+1}}\cdots e^{a_D}.
\end{eqnarray}
Of these, $\mathcal{P}_{2k}$ is the Pontryagin form and $\mathcal{E}_{2n}$is the Euler form. Their integrals are topological invariants in $4k$ and $2n$ dimensions, respectively. If integrated on compact manifolds without boundary, 
\begin{equation}
\Omega_n \int_{M^{2n}} \mathcal{E}_{2n} \in \mathbb{Z}, \quad \quad \tilde{\Omega}_k \int_{M^{4k}} \mathcal{P}_{4k} \in \mathbb{Z}.
\end{equation}
Thus, in every even dimension $D=2n$ there is a topological invariant of the Euler family. If the dimension is a multiple of four, there are invariants of the Pontryagin family as well, of the form
\begin{equation}
\mathcal{P}_{D}=\mathcal{P}_{2k_1}\mathcal{P}_{2k_2}\cdots \mathcal{P}_{2k_r},
\end{equation}
where $4(k_1 +k_2 +\cdots+ k_r)=D$. While there is only one Euler density, the number of Pontryagin invariants grows with the number of partition of the number $D/4$.

\subsubsection{Lovelock theory}      
If torsion is set to zero, the invariants $\tau_k$, $\zeta_k$ clearly vanish, but also $\upsilon_k$ must vanish, since the contraction $R^a{}_b e^b$ it equals the covariant derivative of the torsion. Hence, we are led to the following

\smallskip{\bf Theorem }[Lovelock, 1970 \cite{Lovelock} and Zumino, 1986 \cite{Zumino}]: In the absence of torsion, the most general action for gravity $I[e,\omega]$ invariant, under Lorentz transformations that does not involve explicitly the metric, is of the form
\begin{equation}
I_D[e,\omega]=\kappa \int_{M} \sum_{p=0}^{[D/2]}a _{p}L^D_p \label{LL-action}
\end{equation}
where $a_p$ are arbitrary constants, and $L^D_p $ is given by
\begin{equation}
L^D_p=\epsilon_{a_{1}\cdots a_{D}}R^{a_{1}a_{2}}\cdots R^{a_{2p-1}a_{2p}}e^{a_{2p+1}}\cdots e^{a_{D}}. \;\; \blacksquare
\label{Lovlag}
\end{equation}

The Lovelock series is an arbitrary linear combination where each term $L^D_p$ is the continuation to dimension $D$ of all the lower-dimensional Euler forms. In even dimensions, the last term in the series is the Euler form of the corresponding dimension, $L^D_{D/2}=\mathcal{E}_D$. Let us examine a few examples.\\

$\bullet$ $D=2$: The Lovelock Lagrangian reduces to 2 terms, the $2$-dimensional Euler form and the spacetime volume (area),
\begin{eqnarray}
I_{2} &=&\kappa \int_{M} \epsilon_{ab} [a_{1} R^{ab} + a_{0} e^a e^b] \nonumber \\
          &=&\kappa \int_M \sqrt{|g|}\left( a_{1}R + 2 a_{0}\right) d^{2}x  \label{2DGrav} \\
          &=&\kappa a_{1}\cdot \mathcal{E}_{2}+ 2\kappa a _{0}\cdot V_{2}. \nonumber
\end{eqnarray}
This action has as a local extremum for $V_2=0$, which reflects the fact that, unless matter is included, $I_{2}$ does not make a very interesting dynamical theory for the geometry. If the manifold $M$ has Euclidean metric and a prescribed boundary, the first term picks up a boundary term and the action is extremized by a minimal surface, like a soap bubble, the famous Plateau problem.

$\bullet$ $D=3$ and $D=4$: The action reduces to the Hilbert action plus a volume term, the cosmological constant. In four dimensions, the action admits, in addition, the four dimensional Euler invariant $\mathcal{E}_{4}$,
\begin{eqnarray}
I_{4} &=&\kappa \int_M \epsilon_{abcd} \left[ a_2 R^{ab}R^{cd}+ a_1 R^{ab}e^c e^d + a_{0}e^a e^b e^ c e^d \right] \nonumber \\
 &=&-\kappa \int_M \sqrt{|g|}\left[a_2\left(R^{\alpha \beta \gamma  \delta }R_{\alpha \beta \gamma \delta}-4R^{\alpha \beta}R_{\alpha \beta} +R^{2}\right) + 2a_1R + 24 a_0 \right] d^4 x \nonumber \\
&=&-\kappa a_{2}\cdot \mathcal{E}_{4}- 2a _1\int_M \sqrt{|g|}R d^4 x  - 24\kappa a_0 \cdot V_{4}. \label{4DGrav}
\end{eqnarray}

$\bullet$ $D=5$: The Euler form $\mathcal{E}_4$, also known as the Gauss-Bonnet density, provides the first nontrivial generalization of Einstein gravity occurring in five dimensions,
\begin{equation}
\epsilon _{abcde}R^{ab}\!R^{cd}e^{e}\!= \sqrt{|g|}\left[R^{\alpha \beta \gamma \delta }R_{\alpha \beta \gamma \delta }-4R^{\alpha \beta }R_{\alpha \beta} +R^{2}\right] d^5 x  \label{G-B}.
\end{equation}

The fact that this term could be added to the Einstein-Hilbert action in five dimensions seems to have been known for many years. This is commonly attributed to Lanczos \cite{Lanczos}, but the original source is unclear. 

\subsubsection{Dynamical content of Lovelock theory}   

The Lovelock theory is the natural generalization of GR when the spacetime dimension is greater than four. In the absence of torsion this theory generically describes the same $D(D-3)/2$ degrees of freedom as the Einstein-Hilbert theory \cite{Te-Z}. The action (\ref{LL-action}) has been identified as describing the only ghost-free\footnote{A Lagrangian containing arbitrarily high derivatives of fields generally leads to ghosts.} effective theory for a spin two field, generated from string theory at low energy \cite{Zwiebach,Zumino}. The unexpected and nontrivial absence of ghosts seems to reflect the fact that in the absence of torsion, the Lovelock action yields at most second order field equations for the metric, so that the propagators behave as $ k^{-2}$, and not as $ k^{-2}+ k^{-4}$, as would be the case in a generic higher derivative theory, involving arbitrary combinations and higher powers of the curvature tensor.

Extremizing the action (\ref{LL-action}) with respect to $e^{a}$ and $\omega^{ab}$, yields
\begin{equation}
\delta I_{D}=\int [\delta e^{a}{\cal E}_{a}+\delta \omega^{ab}{\cal E}_{ab}]=0,
\label{Var-action}
\end{equation}
modulo surface terms. The condition for $I_{D}$ to have an extremum to first order under arbitrary infinitesimal variations is that ${\cal E}_{a}$ and ${\cal E}_{ab}$ vanish:
\begin{equation}
{\cal E}_{a}=\sum_{p=0}^{[\frac{D-1}{2}]} a_p(D-2p){\cal E}_{a}^{(p)}=0, \label{D-curvature}
\end{equation}
and
\begin{equation}
{\cal E}_{ab}=\sum_{p=1}^{[\frac{D-1}{2}]} a_p p(D-2p){\cal E}_{ab}^{(p)}=0. \label{D-torsion}
\end{equation}
where we have defined
\begin{eqnarray}
{\cal E}_{a}^{(p)}&:=&\epsilon_{ab_2 \cdots b_D} R^{b_2 b_3} \cdots R^{b_{2p}b_{2p+1}}e^{b_{2p+2}}\cdots e^{b_D},\hspace{-0.06in}  \label{ELL1}
\\
{\cal E}_{ab}^{(p)}&:=&\epsilon_{aba_3 \cdots a_D} R^{a_3 a_4 } \cdots R^{a_{2p-1}a_{2p}} T^{a_{2p+1}} e^{a_{2p+2}} \cdots e^{a_D}.
\label{ELL2}
\end{eqnarray}
These equations involve only first derivatives of $e^{a}$ and $\omega_{\;b}^{a}$, simply because $d^{2}=0$. If one furthermore assumes --as is usually done-- that the torsion vanishes identically,
\begin{equation}
T^{a}=de^{a}+\omega^a{}_b e^b=0,  \label{0-Torsion}
\end{equation}
then Eq. (\ref{D-torsion}) is automatically satisfied. Moreover, the torsion-free condition can be solved for $\omega$ as a function of the inverse vielbein ($E^{\mu}_a$) and its derivative,
\begin{equation}
\omega^a{}_{b \mu} =-E^{\nu}_b(\partial_{\mu}e^a_{\nu} - \Gamma^{\lambda}_{\mu \nu} e^a_{\lambda}),
\end{equation}
where $\Gamma^{\lambda}_{\mu \nu}$ is symmetric in $\mu \nu$ and can be identified as the Christoffel symbol (torsion-free affine connection). Substituting this expression for the Lorentz connection back into (\ref{ELL1}) yields second order field equations for the metric. 

These equations are identical to the ones obtained from varying the Lovelock action written in terms of the Riemann tensor and the metric,
\[
I_{D}[g]=\! \int_M\!\!\! d^{D}x\sqrt{g}\left[ a' _0 + a'_1 R + a'_2 (R^{\alpha \beta \gamma \delta} R_{\alpha \beta \gamma\delta}- 4R^{\alpha \beta }R_{\alpha \beta} + R^2) + \cdot\! \cdot\! \cdot\! \right].
\]
Now one can understand the remarkable feature of GR, where the field equations for the metric are second order and not fourth order, in spite of the fact that the Lagrangian involves the second derivatives of $g_{\mu \nu}$. This ``miraculous accident" is a consequence of the fact that the action can be written using only wedge products and exterior derivatives of the fields, without using the *-Hodge dual, and the fact that the torsion is assumed to vanish identically. In $f(R)$ theories, for example, this first condition is not respected and they generically give rise to higher order field equations. In the presence of fermionic matter, torsion does not vanish, hence the second condition would not hold and the use of a purely metric formulation would be unwarranted.

The standard, purely metric, form of the action is also called second order formalism, because it yields equations with up to second derivatives of the metric. The fact that the Lagrangian contains second derivatives of $g_{\mu \nu}$ has induced some authors to refer to the Lovelock actions as \textit{higher derivative theories of gravity}, which is incorrect, as already mentioned.

One important feature that makes the behaviour of Lovelock theories very different for $D\leq4$ and for $D> 4$ is that in the former case the field equations (\ref{D-curvature}, \ref{D-torsion}) are linear in the curvature tensor, while in the latter case the equations are generically nonlinear in $R^{ab}$. In particular, while for $D\leq 4$ the equations (\ref{ELL2}) imply the vanishing of torsion, this is no longer true for $D>4$. In fact, the field equations evaluated in some configurations may leave some components of the curvature and torsion tensors completely undetermined. For example, Eq.(\ref{D-torsion}) has the form of a polynomial in $R^{ab}$ times $T^{a}$, and it is possible that the polynomial vanishes identically, imposing no conditions on the torsion tensor. 

However, the configurations for which the equations do not determine $R^{ab}$ and $T^{a}$ form sets of measure zero in the space of geometries. In a generic case, outside of these degenerate configurations, the Lovelock theory has the same number of degrees of freedom as ordinary gravity \cite{Te-Z}. The problem of degeneracy, however, is a major issue in determining the time evolution of certain dynamical systems, usually associated with the splitting of the phase space into causally disconnected regions and irreversible loss of degrees of freedom \cite{STZ}. These features might be associated to a dynamical dimensional reduction in gravitation theories \cite{HTrZ}, and has been shown to survive even at the quantum level \cite{dM-Z}.

\subsubsection{Torsional series}   

Lovelock's theorem assumes equation (\ref{0-Torsion}) to be an identity. This means that $e^a$ and $\omega_{\;b}^a$ are not independent fields, contradicting the assumption that these fields correspond to two independent features of the geometry on equal footing. For $D\leq 4$, equation (\ref{0-Torsion}) coincides with (\ref{ELL2}), so that imposing the torsion-free constraint may be seen as an unnecessary albeit harmless restriction. In fact, for 3 and 4 dimensions, the Lorentz connection can be algebraically obtained from its own field equation and by the implicit function theorem, the first order and the second order actions have the same extrema and define equivalent theories, $I[\omega ,e] =I[\omega (e,\partial e),e]$. 

As can be seen from  (\ref{D-torsion}), the torsion-free condition does not automatically follow from the field equations, and although (\ref{ELL2}) is algebraic in $\omega$, it is impossible to solve for $\omega(e,\partial e)$ globally. Therefore, the second order action is not necessarily equivalent to the first order one in general. It could be that for some choices of coefficients $a_p$, the curvature is such that leaves the torsion completely indeterminate. 

Thus, it is reasonable to consider the generalization of the Lovelock action in which torsion \textit{is not} assumed to vanish identically, adding of all possible Lorentz invariants involving torsion that would vanish if $T^a=0$ \cite{Mardones-z}. This means allowing for combinations of terms included in the first four expressions (\ref{P}-\ref{z}). For example, a possible contribution to the Lagrangian in fifteen dimensions could be $\mathcal{P}_4 \upsilon_1 \tau_0 \zeta_0$. Let's examine other examples:\\
$\bullet$ For $D=3$, there is one torsion term in addition to the Lovelock family,
\begin{equation}
\zeta_0=e^{a}T_{a},  \label{eT}
\end{equation}
$\bullet$ For $D=4$, there are three such terms,
\begin{equation}
\upsilon_1= e^{a}e^{b}R_{ab}, \;\;\; \tau_0=T^{a}T_{a}, \;\; \; \mathcal{P}_4= R^{ab}R_{ab}.\;\;  \label{e2R+T2}
\end{equation}
The integral of the last term in (\ref{e2R+T2}) is a topological invariant (the Pontryagin number), and the linear combination of the other two terms,
\begin{equation}
\mathcal{N}_{4}=T^{a}T_{a}-e^{a}e^{b}R_{ab},  \label{NY}
\end{equation}
known as the Nieh-Yan form, also yields a topological invariant \cite{Nieh-Yan}. The properly normalized integral of (\ref{NY}) over a 4-manifold is an integer, equal to the difference between the Pontryagin classes for $SO(5)$ and $SO(4)$ (or their related groups $SO(n,5-n$ and $SO(m,4-m)$) \cite{ChaZ}. 

To make life even harder, there are some linear combinations of these products which are topological densities, as in (\ref{e2R+T2}). In 8 dimensions there are two Pontryagin forms
\begin{eqnarray}
\mathcal{P}_8 &=&R_{\;a_2}^{a_1} R_{\;a_3}^{a_2} \cdots R_{\;a_1}^{a_4}, \\
(\mathcal{P}_4)^2 &=&(R_{\;b}^a R_{\;a}^b)^2,
\end{eqnarray}
which also occur in the absence of torsion, and there are two generalizations of the Nieh-Yan form,
\begin{eqnarray}
(\mathcal{N}_4)^2 &=& (T^a T_a -e^a e^b R_{ab})^2, \\ 
\mathcal{N}_4 \mathcal{P}_4  &=& (T^a T_a -e^a e^b R_{ab}) (R_{\;d}^c R_{\; c}^d),
\end{eqnarray}
etc. (for details and extensive discussions, see Ref.\cite{Mardones-z}).

\subsubsection{Quasi-invariant Chern-Simons series}   
The Pontryagin classes $\mathcal{P}_{2n}$ defined in (\ref{P}), as well as those  that involve torsion, like the Nieh-Yan forms (\ref{NY}), are all closed forms. Therefore, one can look for a locally defined CS form whose exterior derivative yields the corresponding closed form. These CS forms can also be included as Lagrangian densities in the appropriate dimension.  

The idea is best illustrated with examples. Consider the Pontryagin and the Nieh-Yan forms in four dimensions, $\mathcal{P}_4$ and $\mathcal{N}_4$, respectively. The corresponding CS three-forms are
\begin{eqnarray}
C^{Lor}_3 &=& \omega^a_{\;b} d\omega^b_{\;a} + \frac{2}{3} \omega^a_{\;b} \omega^b_{\;c} \omega^c_{\;a} \\
C^{Tor}_3&=& e^aT_a
\end{eqnarray}
Both these terms are invariant under $SO(2,1)$ (Lorentz invariant in three dimensions), and are related to the four-dimensional Pontryagin and Nieh-Yan forms,
\begin{eqnarray}
dC^{Lor}_3 &=& R^{ab}R_{ab}, \\
dC^{Tor}_3&=& T^{a}T_{a}-e^{a}e^{b}R_{ab}.
\end{eqnarray}

The general recipe is simple. For each Pontryagin form in $4k$ dimensions, there is a CS form which provides a sensible action for gravity in $4k-1$ dimensions. For example, in $D=7$, the Lorentz CS form is
\[
C^{Lor}_7 = Tr[\omega (d\omega)^3 + \frac{8}{5}\omega^3 (d\omega)^2 +\frac{4}{5}\omega^2 (d\omega)\omega (d\omega)+2\omega^5(d\omega) +\frac{4}{7}\omega^7],
\]
where the trace is over the suppressed the Lorentz indices.

Thus, the most general gravity in a given dimension would be a linear combination of Lorentz invariant and quasi-invariant $D$-forms of the three families described above: Lovelock, torsional and Lorentz Chern-Simons forms. While the Lovelock series has a simple systematic rule for any dimension (\ref{LL-action}), no simple recipe is there for the torsional Lagrangians. These look awkward, there is no systematic rule to even say how many terms appear in a given dimension and the number of elementary terms of the families $\upsilon$, $\tau$ and $\zeta$, grows wildly with the dimension\footnote{As it is shown in \cite{Mardones-z}, the number of torsion-dependent terms grows as the partitions of $D/4$, which is given by the Hardy-Ramanujan formula, $p(D/4)\sim \frac{1}{\sqrt{3}D}\exp [\pi \sqrt{D/6}]$.}. This proliferation problem is not purely aesthetic. It is like the cosmological constant problem but for a huge number of indeterminate parameters in the theory and not just one.

\section{Selecting Sensible Theories}  
There is another serious aspect of the proliferation issue: the coefficients in front of each term in the Lagrangian are not only arbitrary but dimensionful. This problem already occurs in 4 dimensions, where Newton's constant and the cosmological constant have dimensions of [length]$^{2}$ and [length]$^{-4}$ respectively. 

The presence of dimensionful parameters leaves little room for optimism in a quantum version of the theory. Dimensionful parameters in the action are potentially dangerous because they are likely to acquire uncontrolled quantum corrections. This is what makes ordinary gravity nonrenormalizable in perturbation theory: In 4 dimensions, Newton's constant has dimensions of [mass]$^{-2}$ in natural units. This means that as the order in perturbation series increases, more powers of momentum will occur in the Feynman graphs, making the ultraviolet divergences increasingly worse. Concurrently, the radiative corrections to these bare parameters require the introduction of infinitely many counterterms into the action to render them finite \cite{tHooft}. But an illness that requires infinite amount of medication is synonym of incurable.

The only safeguard against the threat of uncontrolled divergences in quantum theory is to have some symmetry principle that fixes the values of the parameters in the action, limiting the number of possible counterterms that could be added to the Lagrangian. Obviously, a symmetry endowed with such a high responsibility should be a \textit{bona fide} quantum symmetry, and not just an approximate feature of its effective classical descendent. A symmetry that is only present in the classical limit but is not a feature of the quantum theory is said to be anomalous. This means that if one conceives the quantum theory as the result of successive quantum corrections to the classical theory, these corrections ``break" the symmetry. An anomalous symmetry is an artefact of the classical limit, that does not correspond to a true symmetry of the microscopic world.  

If a non anomalous symmetry fixes the values of the parameters in the action, this symmetry will protect those values under renormalization. A good indication that this might happen would be if all the coupling constants are dimensionless and could be absorbed in the fields, as in Yang-Mills theory. As shown below, in odd dimensions there is a unique choice of coefficients in the Lovelock action that gives the theory with an enlarged gauge symmetry. This action has no dimensionful parameters and can be seen to depend on a unique (dimensionless) coefficient ($\kappa$), analogous to Newton's constant. This coefficient can be shown to be quantized by an argument similar to the one that yields Dirac's quantization of the product of magnetic and electric charge \cite{QuantumG}. All these miraculous properties can be traced back to the fact that the particular choice of coefficients in that Lagrangian turns it into a CS form for the enlarged gauge symmetry.

\subsection{Extending the Lorentz group}    

The coefficients $\alpha _{p}$ in the Lovelock action (\ref{LL-action}) have dimensions $l^{D-2p}$. This is because the canonical dimension of the vielbein is $[e^{a}]=l$, while the Lorentz connection has dimensions $[\omega^{ab}]=$ $l^{0}$, as a true gauge field. This reflects the fact that gravity is naturally a gauge theory for the Lorentz group, where $e^a$ plays the role of a matter field, \textit{not} a connection field but a vector under Lorentz transformations.

\subsubsection{Poincar\'{e} group}    

Three-dimensional gravity, where $e^{a}$ can play the role of a connection, is an important exception to the previous statement. This is in part thanks to the coincidence in three dimensions that allows to regard a vector as a connection for the Lorentz group, $\hat{ \omega}^a=\frac{1}{2}\epsilon ^{abc} \omega_{bc}$. Consider the Einstein-Hilbert Lagrangian in three dimensions
\begin{equation}
L_{3}=\epsilon_{abc}R^{ab}e^{c}.  \label{D=3}
\end{equation}
Under an infinitesimal Lorentz transformation with parameter $\lambda_{\,\;b}^{a}$, the Lorentz connection transforms as
\begin{eqnarray}
\delta \omega_{\,\;b}^{a} &=&D\lambda_{\,\;b}^{a} \label{delta w}\\
&=&d\lambda_{\,\;b}^{a} + \omega _{\,\;c}^{a}\lambda_{\,\; b}^{c} -
\omega_{\,\;b}^{c}\lambda _{\,\;c}^{a}, \nonumber
\end{eqnarray}
while $e^{c}$, $R^{ab}$ and $\epsilon_{abc}$ transform as tensors,
\begin{eqnarray*}
\delta e^{a} &=&-\lambda _{\,\;c}^{a}e^{c} \\ \delta R_{\,}^{ab} &=&-(\lambda_{\,\;c}^{a}R^{cb}+\lambda_{\,\; c}^{b}R^{ac}), \\
\delta \epsilon_{abc} &=&-(\lambda_{\,\;a}^{d}\epsilon_{dbc}+\lambda_{\,\;b}^{d}\epsilon_{adc}+\lambda_{\,\;c}^{d}\epsilon_{abd})\equiv0.
\end{eqnarray*}
Combining these relations, the Lorentz invariance of $L_{3}$ can be directly checked. What is unexpected is that the action defined by (\ref{D=3}) is also invariant under the group of local translations in the three dimensional tangent space. For this additional symmetry  $e^{a}$ transforms as a gauge connection for the translation group.\footnote{This translational invariance in the tangent space is not to be confused with the local translations in the base manifold mentioned in Sect.1.3.1, Eq(\ref{diffeos}).} In fact, if the vielbein transforms under ``local translations'' in tangent space, parametrized by $\lambda^{a}$ as 
\begin{eqnarray}
\delta e^{a} &=&D\lambda ^{a}  \nonumber \\
&=&d\lambda ^{a}+\omega _{\,\;b}^{a}\lambda^{b},  \label{LocalTrans-e}
\end{eqnarray}
while the Lorentz connection remains unchanged,
\begin{equation}
\delta \omega^{ab} = 0, \label{LocalTrans-w}
\end{equation}
then, the Lagrangian $L_{3}$ changes by a total derivative,
\begin{equation}
\delta L_{3}=d[\epsilon_{abc}R^{ab}\lambda^{c}],  \label{dL3}
\end{equation}
which can be dropped from the action, under the assumption of standard boundary conditions. This means that in three dimensions ordinary gravity is gauge invariant under the whole Poincar\'{e} group. (This can be shown using the infinitesimal transformations $\delta e$ and $\delta \omega$ to compute the commutators of the second variations, obtaining the Lie algebra of the Poincar\'{e} group.)

\subsubsection{(Anti-)de Sitter group}    

In the presence of a cosmological constant $\Lambda =\mp l^{-2}$ it is also possible to extend the local Lorentz symmetry. In this case, however, the invariance of the appropriate tangent space is not the local Poincar\'e  symmetry, but the local (anti)-de Sitter group. The point is that different spaces $T^{*}M$ can be chosen as tangents to a given manifold $M$, provided they are diffeomorphic to the open neighbourhoods of $M$. However, a useful choice of tangent space corresponds to the covering space of a vacuum solution of the Einstein equations. In the previous case, flat space was singled out because it is the maximally symmetric solution of the Einstein equations. If $\Lambda \neq 0$,  flat spacetime is no longer a solution of the Einstein equations, but the de Sitter or anti-de Sitter  space,  for $\Lambda > 0$ or $\Lambda < 0$, respectively.

The three-dimensional Lagrangian in (\ref{LL-action}) reads
\begin{equation}
L_{3}^{AdS}=\epsilon _{abc}(R^{ab}e^{c}\pm \frac{1}{3l^{2}}e^{a}e^{b}e^{c}),
\label{L3AdS}
\end{equation}
and the action is invariant --modulo surface terms-- under the infinitesimal transformations,
\begin{eqnarray}
\delta \omega _{\,\;}^{ab} &=&d\lambda _{\;}^{ab}+ \omega_{\,\;c}^{a} \lambda^{cb}+\omega_{\,\;c}^{b} \lambda_{\;}^{ac} 
\pm [e^{a}\lambda ^{b}-\lambda ^{a}e^{b}]l^{-2} \label{dw-AdS} \\ 
\delta e^{a} &= &d\lambda^{a}+\omega _{\,\;b}^{a}\lambda ^{b} - \lambda_{\;b}^ae^b. \label{de-AdS}
\end{eqnarray}
These transformations can be cast in a more suggestive way as
\begin{eqnarray*}
\delta \left[
\begin{array}{cc}
\omega _{\,\;}^{ab} & e^a l^{-l} \\
-e^b l^{-l} & 0
\end{array}
\right] &=& d\left[
\begin{array}{cc}
\lambda _{\;}^{ab} & \lambda^a l^{-l} \\
-\lambda^b l^{-l} & 0
\end{array}
\right] \\
&& +\left[
\begin{array}{cc}
\omega _{\;c}^{a} & e^a l^{-l} \\
-e_{c}l^{-l} & 0
\end{array}
\right] \left[
\begin{array}{cc}
\lambda ^{cb} & \lambda^{c}l^{-1} \\
\pm\lambda^{b}l^{-1} & 0
\end{array}
\right] \\
&&- \left[
\begin{array}{cc}
\lambda ^{ac} & \lambda^{a} l^{-1} \\
-\lambda^{c} l^{-1}& 0
\end{array}
\right] \left[
\begin{array}{cc}
\omega _{c}^{\;b} &  e_c l^{-1} \\
\pm e^bl^{-1} & 0
\end{array}
\right] .
\end{eqnarray*}
This can also be written as
\[
\delta W^{AB}=d\Lambda_{\;}^{AB}+W_{\,\;C}^{A}\Lambda ^{CB}- \Lambda^{AC}W_{C}^{\;B} ,
\]
where the 1-form $W^{AB}$ and the 0-form $\Lambda^{AB}$ stand for the combinations
\begin{eqnarray}
W_{\;}^{AB} &=&\left[
\begin{array}{cc}
\omega _{\,\;}^{ab} & e^{a}l^{-1} \\
-e^{b}l^{-1} & 0
\end{array}
\right]  \label{baticonn} \\
\Lambda ^{AB} &=&\left[
\begin{array}{cc}
\lambda_{\;}^{ab} & \lambda^{a}l^{-1} \\
-\lambda^{b}l^{-1} & 0
\end{array}
\right] ,  \label{batiparam}
\end{eqnarray}
(here $a,b,..=1,2,..D,$ while $A,B,...=1,2,..,D+1$). Clearly, $W_{\; }^{AB}$ transforms as a connection and $\Lambda^{AB}$ can be identified as the
infinitesimal transformation parameters, but for which group? Since $\Lambda^{AB}=-\Lambda^{BA}$, this indicates that the group is one that leaves invariant a symmetric, real bilinear form, so it must be a group in the $SO(r,s)$ family. The signs ($\pm $) in the transformation above can be traced back to the sign of the cosmological constant. It is easy to check that this structure fits well if indices are raised and lowered with the metric
\begin{equation}
\Pi^{AB}=\left[
\begin{array}{cc}
\eta _{\,\;}^{ab} & 0 \\
0 & \mp 1
\end{array}
\right] ,  \label{tangentAdS}
\end{equation}
so that, for example, $W^{AB}=\Pi^{BC}W^A_{\,\;C}$. Then, the covariant derivative in the connection $W$ of this metric vanishes identically,
\begin{equation}
D_{W}\Pi^{AB}=d\Pi^{AB}+W^{A}_{\,\;C}\Pi^{CB}+W^{B}_{\,\; C}\Pi ^{AC}=0.
\end{equation}
Since $\Pi^{AB}$ is constant, this last expression implies $W^{AB}+W^{BA}=0$, in exact analogy with what happens with the Lorentz connection,
$\omega^{ab}+\omega ^{ba}=0$, where $\omega ^{ab}\equiv\eta ^{bc}\omega_{\;c}^{a}$. Indeed, this is a very awkward way to discover that the 1-form $W_{\;}^{AB}$ is actually a connection for the group which leaves invariant the metric $\Pi^{AB}$. Here the two signs in $\Pi^{AB}$ correspond to the de Sitter ($+$) and anti-de Sitter\ ($-$) groups, respectively.

What we have found here is an explicit way to immerse the three-dimensional Lorentz group into a larger symmetry group, in which the vielbein and the Lorentz connection have been incorporated on equal footing as components of a larger (A)dS connection. The Poincar\'{e} symmetry is obtained in the limit $l\rightarrow \infty $ ($\lambda\rightarrow 0 $). In that case, instead of (\ref{dw-AdS}, \ref{de-AdS}) one has
\begin{eqnarray}
\delta \omega _{\,\;}^{ab} &=&d\lambda _{\;}^{ab}+ \omega_{\,\;c}^{a} \lambda^{cb}+\omega_{\,\;c}^{b}\lambda_{\; }^{ac}  \label{dw-Poinc} \\ \delta e^{a} &=&d\lambda ^{a}+\omega _{\,\;b}^{a}\lambda ^{b} - \lambda _{\;b}^{a}e^{b}. \label{de-Poinc}
\end{eqnarray}
The vanishing cosmological constant limit is actually a deformation of the (A)dS algebra analogous to the deformation that yields the Galileo group from the Poincar\'e symmetry in the limit of infinite speed of light ($c\rightarrow \infty$). These deformations are examples of what is known as a In\"on\"u-Wigner contraction \cite{Gilmore,Inonu}. The procedure starts from a semisimple Lie algebra and some generators are rescaled by a parameter ($l$ or $\lambda$ in the above example). Then, in the limit where the parameter is taken to zero (or infinity),  a new (not semisimple) algebra is obtained.  For the Poincar\'e group which is the familiar symmetry of Minkowski space, the representation in terms of $W$ becomes inadequate because the metric $\Pi^{AB}$
should be replaced by the degenerate (noninvertible) metric of the Poincar\'{e} group,
\begin{equation}
\Pi_0^{AB}=\left[
\begin{array}{cc}
\:\eta _{\,\;}^{ab} & 0 \\
0 & 0
\end{array}
\right] ,  \label{tangentPoinc}
\end{equation}
and is no longer clear how to raise and lower indices. Nevertheless, the Lagrangian (\ref{L3AdS}) in the limit $l\rightarrow \infty $ takes the usual
Einstein Hilbert form with vanishing cosmological constant,
\begin{equation}
L_{3}^{EH}=\epsilon_{abc}R^{ab}e^{c}, \label{L3EH}
\end{equation}
which is invariant under (\ref{de-Poinc}). 

As Witten showed, General Relativity in three spacetime dimensions is a renormalizable quantum system \cite{Witten}. It is strongly suggestive that precisely in 2+1 dimensions GR is also a gauge theory on a fibre bundle. It could be thought that the exact solvability miracle is due to the absence of propagating degrees of freedom in three-dimensional gravity, but the final power-counting argument of renormalizability rests on the fibre bundle structure of the Chern-Simons system and doesn't seem to depend on the absence of propagating degrees of freedom. In what follows we will generalize the gauge invariance of three-dimensional gravity to higher dimensions.

\subsection{More Dimensions}     

Everything that has been said about embedding the Lorentz group into the (A)dS group for $D=3$, starting at equation (\ref{dw-AdS}), can be generalized for any $D$. In fact, it is always possible to embed the $D$-dimensional Lorentz group into the de-Sitter, or anti-de Sitter groups,
\begin{equation}
SO(D-1,1)\hookrightarrow \left\{
\begin{array}{cc}
SO(D,1), & \Pi ^{AB}=\rm{diag }(\eta _{\,\;}^{ab},+1) \\
SO(D-1,2), & \Pi ^{AB}=\rm{diag }(\eta _{\,\;}^{ab},-1)
\end{array}
.\right.  \label{embeddingAdS}
\end{equation}
as well as into their Poincar\'{e} limit,
\begin{equation}
SO(D-1,1)\hookrightarrow ISO(D-1,1).  \label{embeddingPoinc}
\end{equation}

The question naturally arises,  are there gravity actions in dimensions $\geq 3$ which are also invariant, not just under the Lorentz group, but under some of its extensions, $SO(D,1)$, $SO(D-1,2)$, $ISO(D-1,1)$? As we will see now, the answer to this question is affirmative in odd dimensions: There exist gravity actions for every $D=2n-1$, invariant under local $SO(2n-2,2)$, $SO(2n-1,1)$ or $ISO(2n-2,1)$ transformations, where the vielbein and the Lorentz connection combine to form the connection of the larger group. In even dimensions, in contrast, this cannot be done.

Why is it possible in three dimensions to enlarge the symmetry from local $SO(2,1)$ to local $SO(3,1)$, $SO(2,2)$ and $ISO(2,1)$? What happens if one tries to do this in four or more dimensions? Let us start with the Poincar\'{e} group and the Hilbert action for $D=4$,
\begin{equation}
L_{4}=\epsilon _{abcd}R^{ab}e^{c}e^{d}.  \label{D=4}
\end{equation}
Why is this not invariant under local translations $\delta e^{a}= d\lambda^{a}+\omega _{\,\;b}^{a}\lambda ^{b}$? A simple calculation yields
\begin{eqnarray}
\delta L_{4} &=&2\epsilon _{abcd}R^{ab}e^{c}\delta e^{d}  \nonumber \\
&=&d(2\epsilon _{abcd}R^{ab}e^{c}\lambda ^{d})-2\epsilon_{abcd}R^{ab}T^{c}\lambda ^{d}.  \label{delta4}
\end{eqnarray}
The first term in the r.h.s. of (\ref{delta4}) is a total derivative and therefore gives a surface contribution to the action. The last term, however, need not vanish, unless one imposes the field equation $T^{a}=0$. But this means that the invariance of the action only occurs \textit{on shell}. Now, ``on shell symmetries" are not real symmetries and they probably don't survive quantization because quantum mechanics does not respect equations of motion.

On the other hand, the miracle in three dimensions occurred because the Lagrangian (\ref{L3EH}) is linear in $e$. In fact, Lagrangians of the form
\begin{equation}
L_{2n+1}=\epsilon_{a_{1}\cdots a_{2n+1}}R^{a_{1}a_{2}}\cdots R^{a_{2n-1}a_{2n}}e^{a_{2n+1}}, \label{odd-D-Poinc}
\end{equation}
--which are only defined in odd dimensions--, are also invariant under local Poincar\'{e} transformations (\ref{dw-Poinc}, \ref{de-Poinc}), as can be easily checked. Since the Poincar\'{e} group is a limit of (A)dS, it seems likely that there should exist a Lagrangian in odd dimensions, invariant under local (A)dS transformations, whose limit for vanishing cosmological constant ($l\rightarrow \infty$) is (\ref{odd-D-Poinc}). One way to find out what that Lagrangian might be, one could take the most general Lovelock Lagrangian and select the coefficients by requiring invariance under (\ref{dw-AdS}, \ref{de-AdS}). This is a long and tedious but sure route. An alternative approach is to try to understand why it is that in three dimensions the gravitational Lagrangian with cosmological constant (\ref{L3AdS}) is invariant under the (A)dS group.

Let us consider the three-dimensional case first. If we take seriously the notion that $W^{AB}$ is a connection, then the associated curvature is
\[
F^{AB}=dW^{AB}+W_{\;C}^{A}W^{CB},
\]
where $W^{AB}$ is defined in (\ref{baticonn}). Then, it is easy to prove that
\begin{equation}
F_{\;}^{AB}=\left[
\begin{array}{cc} R_{\,\;}^{ab}\pm l^{-2}e^{a}e^{b} & l^{-1}T^{a} \\
-l^{-1}T^{b} & 0
\end{array}
\right] .  \label{baticurvature}
\end{equation}
where $a,b$ run from 1 to 3 and $A,B$ from 1 to 4. Since the (A)dS group has an invariant tensor $\epsilon_{ABCD}$, one can construct the 4-form invariant
\begin{equation}
E_{4}=\epsilon _{ABCD}F_{\;}^{AB}F_{\;}^{CD}.  \label{E=F2}
\end{equation}
This is invariant under the (A)dS group and is readily recognized, up to a constant, as the Euler form\footnote{This identification is formal, since the differential forms that appear here are defined in three dimensions, but they can be naturally defined to four dimensions by simply extending the range of coordinate indices. This implies that one is considering the three-dimensional manifold as embedded in four dimensions.} for a four-dimensional manifold whose tangent space is not Minkowski, but has the metric $\Pi ^{AB}= $diag $(\eta _{\,\;}^{ab},\mp 1)$. The form $E_{4}=\#\mathcal{E}_{4}$ can also be written explicitly in terms of $R^{ab}$, $T^{a}$, and $e^{a}$,
\begin{eqnarray}
E_{4} &=& 4\epsilon _{abc}(R_{\,\;}^{ab}\pm l^{-2}e^{a}e^{b})l^{-1}T^{a} \label{E4} \\
&=&\frac{4}{l}d\left[ \epsilon _{abc}\left( R_{\,\;}^{ab}\pm \frac{1}{3l^{2}}e^{a}e^{b}\right) e^{c}\right] ,  \nonumber
\end{eqnarray}
which is, up to constant factors, the exterior derivative of the three-dimensional Lagrangian (\ref{L3AdS}),
\begin{equation}
\mathcal{E}_{4}=dL_{3}^{AdS}.  \label{E4=dL3}
\end{equation}

This is the key point: the l.h.s. of (\ref{E4=dL3}) is invariant under local (A)dS$_3$ by construction. Therefore, the same must be true of the r.h.s.,
\[
\delta \left( dL_{3}^{AdS}\right) =0.
\]
Since the variation is a linear operation, it commutes with the derivative
\[
d\left( \delta L_{3}^{AdS}\right) =0,
\]
which in turn means, by Poincar\'{e}'s Lemma \cite{Spivak} that, locally, $\delta L_{3}^{AdS}= d(something)$. This explains why the action is (A)dS invariant up to surface terms, which is exactly what we found for the variation, [see, (\ref{dL3})]. The fact that three dimensional gravity can be written in this way was observed many years ago in Refs. \cite{Achucarro-Townsend,Witten}.

The key to generalize the (A)dS Lagrangian from $3$ to $2n-1$ dimensions is now clear\footnote{The construction we outline here was discussed in \cite{Chamseddine,Muller-Hoissen}, and also in \cite{JJG,BTZ94}.}:

$\bullet$ First, generalize the Euler density (\ref{E=F2}) to a $2n$-form,
\begin{equation}
E_{2n}=\epsilon _{A_{1}\cdot \cdot \cdot A_{2n}}F^{A_{1}A_{2}}\cdot \cdot \cdot
F^{A_{2n-1}A_{2n}}.  \label{E2n=Fn}
\end{equation}

$\bullet$ Second, express $E_{2n}$ explicitly in terms of $R^{ab}$, $T^{a}$ and $e^{a}$ using (\ref{baticurvature}).

$\bullet$ Write this as the exterior derivative of a $(2n-1)$-form $L_{2n-1}$.

$\bullet$ $L_{2n-1}$ can be used as a Lagrangian in $(2n-1)$ dimensions.

This procedure directly yields the $(2n-1)$-dimensional (A)dS-invariant Lagrangian as
\begin{equation}
L_{2n-1}^{(A)dS}=\sum_{p=0}^{n-1}\bar{\alpha}_{p}L^{2n-1}_p,
\label{(A)dS2n+1}
\end{equation}
where $L^{D}_p$ is given by (\ref{Lovlag}). This is a particular case of a Lovelock Lagrangian in which all the coefficients $\bar{\alpha}_{p}$ have been fixed to take the values
\begin{equation}
\smallskip \bar{\alpha}_{p}=\kappa \cdot \frac{(\pm 1)^{p+1}l^{2p-D}}{(D-2p)}
\left(
\begin{array}{c}
n-1 \\
p
\end{array}
\right) ,\;p=1,2,...,n-1=\frac{D-1}{2},  \label{alphasCS}
\end{equation}
where $1\leq p \leq n-1=(D-1)/2$, and $\kappa$ is an arbitrary dimensionless constant. 

Another interesting exercise is to show that, for AdS, the action (\ref{(A)dS2n+1}) can also be written as \cite{MOTZ1}
\begin{equation}
I_{2n-1}=\frac{\kappa}{l} \int\limits_{M}\int\limits_{0}^{1}dt\;\epsilon_{a_1\cdots a_{2n-1}} R_{t}^{a_1 a_2}\cdots R_{t}^{a_{2n-3} a_{2n-2}} e^{a_{2n-1}} \;, \label{t-AdSAction}
\end{equation}
where we have defined $R_{t}^{ab}:= R^{ab} +(t^2/l^2) e^{a}e^{b}$.

\textbf{Example:} In five dimensions, the (A)dS Lagrangian reads 
\begin{equation}
L_5^{(A)dS}=\frac{\kappa}{l}\epsilon _{abcde}\left[ e^a R^{bc} R^{de} \pm \frac{2}{3l^2} e^a e^b e^c R^{de}
+\frac{1}{5l^4}e^a e^b e^c e^d e^e \right] . \label{(A)dS5}
\end{equation}
The parameter $l$ is a length scale --the Planck length-- and cannot be fixed by theoretical considerations. Actually, $l$ only appears in the combination
\[
\tilde{e}^{a}=l^{-1} e^a,
\]
that could be considered as the ``true'' dynamical field, as is the natural thing to do if one uses $W^{AB}$ instead of ($\omega^{ab}, e^a$). In fact, the Lagrangian (\ref{(A)dS2n+1}) can also be written in terms of $W^{AB}$ as
\[
L_{2n-1}^{(A)dS}=\kappa \cdot \epsilon _{A_1 \cdots A_{2n}}\left[ W(dW)^{n-1}+a_3 W^3 (dW)^{n-2}+\cdots a_{2n-1}W^{2n-1}\right] ,  \label{(A)dS2n+1'}
\]
where all indices are contracted appropriately and the coefficients $a_3$, $\cdots a_{2n-1}$ are dimensionless rational numbers fixed by the condition $dL_{2n-1}^{(A)dS}= \mathcal{E}_{2n}$.

\subsection{Generic Chern-Simons forms}     

The construction outlined above is not restricted to the Euler form, but applies to any gauge invariant of similar nature, generally known as characteristic classes, like the Pontryagin or Chern classes. Their corresponding CS forms were studied first in the context of abelian and nonabelian gauge theories (see, e. g., \cite{Jackiw,Nakahara}). Tables 1 and 2 give examples of CS forms which define Lagrangians in three and seven dimensions, and their corresponding topological invariants,

\begin{center}
\begin{tabular}{|l|l|l|}
\hline 
$D=3$ CS Lagrangians & Characteristic classes & Groups \\
\hline 
\hline 
$L_3^{(A)dS} =\epsilon_{abc}(R^{ab}\pm\frac{e^a e^b}{3l^2})e^c$ & $\mathcal{E}_4=\epsilon_{abc}(R^{ab}\pm\frac{e^a e^b}{l^2}) T^c $ & $SO(4)^{(\dag)}$\\ 
\hline  
$L_3^{Lor} =\omega^a{}_b d\omega^b{}_a +\frac{2}{3} \omega^a{}_b \omega^b{}_c\omega^c{}_a$ &
$\mathcal{P}_4^{Lor} =R^a{}_bR^b{}_a$ & $SO(2,1)$ \\ 
\hline
$L_3^{Tor} = e^aT_a$ & $\mathcal{N}_4 =T^aT_a-e^a e^b R_{ab}$ &$SO(2,1)$ \\ 
\hline 
$ L_{3}^{U(1)}= AdA$ & $\mathcal{P}_4^{U(1)} =FF$ & $ U(1)$ \\
\hline 
$ L_3^{SU(N)} =Tr[{\bf A}d{\bf A+}\frac{2}{3}{\bf AAA]}$ & $\mathcal{P}_4^{SU(4)} =Tr[{\bf FF}]$ & $ SU(N)$ \\
\hline
\end{tabular}
\end{center}
\textbf{Table 1:} Three-dimensional gravitational CS Lagrangians, their related characteristic classes and the corresponding gauge groups.\\
$^{\dag}$ {\small Either this or any of its cousins, $SO(3,1)$, $SO(2,2)$}.\\

Here $R$, $F$, and ${\bf F}$ are the curvatures of the Lorentz, the electromagnetic, and the Yang-Mills ($SU(N)$) connections $\omega _{\;b}^{a}$, $A$ and {\bf A}, respectively; $T$ is the torsion; $\mathcal{E}_4$ and $\mathcal{P}_4$ are the Euler and the Pontryagin densities for the Lorentz group \cite{Eguchi-Gilkey-Hanson}, and $\mathcal{N}_4$ is the Nieh-Yan invariant \cite{Nieh-Yan}. The Lagrangians are invariant (up to total derivatives) under the corresponding gauge groups.

\subsection{Torsional Chern-Simons forms}     

So far we have not included torsion in the CS Lagrangian, but as we see in the table above, it is also possible to construct CS forms that include torsion. All the CS forms above are Lorentz invariant (up to an exact form), but there is a linear combination of the second and third which is invariant under the (A)dS group. This is the so-called exotic gravity \cite{Witten},
\begin{equation}
L_{3}^{Exotic}=L_{3}^{Lor} \pm \frac{2}{l^{2}}L_{3}^{Tor}.  \label{3d-exotic}
\end{equation}
As can be shown directly by taking its exterior derivative, this is invariant under (A)dS:
\begin{eqnarray*}
dL_{3}^{Exotic} &=&R_{\;b}^{a}R_{\;a}^{b}\pm \frac{2}{l^{2}} \left( T^{a}T_{a}-e^{a}e^{b}R_{ab}\right) \\ &=&F_{\;B}^{A}F_{\;A}^{B}.
\end{eqnarray*}
This exotic Lagrangian has the curious property of giving exactly the same field equations as the standard $dL_3^{AdS}$, but interchanged: varying with respect to $e^{a}$ one gives the equation for $\omega ^{ab}$ of the other, and vice-versa.

In five dimensions, the only Lorentz invariant that can be formed using $T^a$ is $R^{ab} T_a e_b$, which is a total derivative, and therefore there no new Lagrangians involving torsion in this case. In seven dimensions there are two Lorentz-Pontryagin, and one torsional CS forms,
\begin{center}
\begin{tabular}{|l|l|}
\hline 
$D=7$ CS Lagrangians & Characteristic classes  \\ 
\hline 
\hline
$L_7^{Lor}=\omega(d\omega)^3 + \frac{8}{5}\omega^3 (d\omega)^2 + \cdots +\frac{4}{7}\omega^7$ & $R^a{}_bR^b{}_c R^c{}_d R^d{}_a$ \\
\hline 
$L_7^I=(\omega^a{}_b d\omega^b{}_a +\frac{2}{3}\omega^a{}_b \omega^b{}_c \omega^c{}_a) R^a{}_b R^b{}_a$ & $(R^a{}_b R^b{}_a)^2$ \\ 
\hline 
$L_7^{II} = (e^a T_a) (R^a{}_b R^b{}_a)$ & $(T^a T_a -e^a e^b R_{ab})R^c{}_d R^d{}_c$
\\ \hline
\end{tabular}
\end{center}
\textbf{Table 2:} Seven-dimensional gravitational CS Lagrangians, their related characteristic classes and the corresponding gauge groups.\\

There exist no CS forms in even dimensions for the simple reason that there are no characteristic classes in odd dimensions. The characteristic classes, like the Euler and the Pontryagin (or Chern-Weil) classes are exterior products of curvature two-forms and therefore are forms of even degree. The idea of characteristic class is one of the unifying concepts in mathematics that connects algebraic topology, differential geometry and algebraic geometry. The theory of characteristic classes explains mathematically why it is not always possible to perform a gauge transformation that makes the connection vanish everywhere even if it is locally pure gauge. The nonvanishing value of a topological invariant signals an obstruction to the existence of a gauge transformation that trivializes the connection globally.

There are basically two types of invariants relevant for a Lorentz invariant theory in an even-dimensional manifold:

$\bullet$ The Euler class, associated with the $O(D-n,n)$ groups. In two dimensions, the Euler number is related to the genus ($g$) of the surface, $\chi = 2-2g$.

$\bullet$ The Pontryagin class, associated with any classical semisimple group $G$. It counts the difference between self dual and anti-self dual gauge connections that are admitted in a given manifold.

The Nieh-Yan invariants correspond to the difference between Pontryagin classes for $SO(D+1)$ and $SO(D)$ in $D$ dimensions \cite{ChaZ}.

As there are no similar invariants in odd dimensions, there are no CS actions for gravity for even $D$, invariant under the (anti-) de Sitter or Poincar\'e groups. In this light, it is fairly obvious that although ordinary Einstein-Hilbert gravity can be given a fibre bundle structure for the Lorentz group, this structure cannot be extended to include local translational invariance.

\subsubsection{Quantization of the gravitation constant} 

The only free parameter in a Chern-Simons action is a multiplicative global coefficient, $\kappa$.  Consider a simply connected, compact $2n-1$ dimensional manifold $M$ whose geometry is determined by an Euler-CS Lagrangian. Suppose $M$ to be the boundary of a $2n$-dimensional compact orientable manifold $\Omega$. Then the action for the geometry of $M$ can be expressed as the integral of the Euler density $\mathcal{E}_{2n}$ over $\Omega$, multiplied by $\kappa$. But since there can be many different manifolds with the same boundary $M$, the integral over $\Omega $ should give the same physical predictions as that over a different manifold, $\Omega'$. In order for this change to leave the path integral unchanged, a minimal requirement would be
\begin{equation}
\kappa \left[ \int_{\Omega }\mathcal{E}_{2n}-\int_{\Omega^{\prime }}\mathcal{E}_{2n}\right] =2n\pi \hbar .  \label{quantum}
\end{equation}
The quantity in brackets --with the appropriate normalization-- is the Euler number of the manifold obtained by gluing $\Omega $ and $\Omega^{\prime}$ along $M$ in the right way to produce an orientable manifold, $\chi [\Omega \cup \Omega']$. This integral can take an arbitrary integer value and hence $\kappa$ must be quantized,
\[
\kappa =nh,
\]
where $h$ is\ Planck's constant \cite{QuantumG}.

\subsubsection{Born-Infeld gravity}  

The closest one can get to a CS theory in even dimensions is with the so-called Born-Infeld ({\bf BI}) theories \cite{JJG,BTZ94,Tr-Z}. The BI Lagrangian is obtained by a particular choice of the $\alpha _{p}$'s in the Lovelock series, so that the Lagrangian takes the form
\begin{equation}
L_{2n}^{BI}=\epsilon _{a_{1}\cdot \cdot \cdot a_{2n}}\bar{R}^{a_{1}a_{2}}\cdots \bar{R}^{a_{2n-1}a_{2n}},  \label{2nBI}
\end{equation}
where $\bar{R}^{ab}$ stands for the combination
\begin{equation}
\bar{R}^{ab}=R^{ab}\pm \frac{1}{l^{2}}e^{a}e^{b}. \label{concircular}
\end{equation}
With this definition it is clear that the Lagrangian (\ref{2nBI}) contains only one free parameter, $l$, which, as explained above, can always be absorbed in a redefinition of the vielbein. This Lagrangian has a number of interesting classical features like simple equations, black hole solutions, cosmological models, etc. \cite{JJG,BTZ94,BHscan}. The simplification comes about because the equations admit a unique maximally symmetric configuration given by $\bar{R}^{ab}=0$, in contrast with the situation when all $\alpha_p$'s are arbitrary. As already mentioned for arbitrary $\alpha_p$'s, the field equations do not determine completely the components of $R^{ab}$ and $T^a$ in general. This is because the high nonlinearity of the equations can give rise to degeneracies. The BI choice is in this respect the best behaved since the degeneracies are restricted to only one value of the radius of curvature ($R^{ab}\pm \frac{1}{l^2} e^a e^b = 0$). At the same time, the BI action has the least number of algebraic constrains required by consistency among the field equations, and it is therefore the one with the simplest dynamical behaviour\cite{Tr-Z}.

\subsection{Finite Action and the Beauty of Gauge Invariance}  

Classical invariances of the action are defined modulo surface terms because they are usually assumed to vanish in the variations, and do not affect the equations in the bulk. This is true for boundary conditions that keep the values of the fields fixed at the boundary: Dirichlet conditions. In a gauge theory, however, it may be more relevant to fix gauge invariant properties at the boundary --like the curvature. These are not precisely Dirichlet boundary conditions, but rather mixed conditions of Dirichlet and Neumann types.

On the other hand, it is also desirable to have an action which has a finite value, when evaluated on a physically observable configuration --e.g., on a classical solution. This is not just for the sake of elegance, it is a necessity if one needs to study the semiclassical thermodynamic properties of the theory. This is particularly true for a theory possessing black holes with interesting thermodynamic features. Moreover, quasi-gauge invariant actions defined on an infinitely extended spacetime are potentially ill defined because, under gauge transformations, the boundary terms could give infinite contributions to the action integral. This would not only cast doubt on the meaning of the action itself, but it would violently contradict the wish to have a gauge invariant action principle.

Changing the Lagrangian by a boundary term may seem innocuous but it is a delicate business. The empirical fact is that adding a total derivative to a Lagrangian in general changes the expression for the conserved Noether charges, and again, possibly by an infinite amount. The conclusion from this discussion is that a regularization principle must be in place in order for the action to be finite on physically interesting configurations, and that assures it remains finite under gauge transformations, and yields well defined conserved charges. 

In \cite{MOTZ1} it is shown that the action has an extremum when the field equations hold, and is finite on classically interesting configuration if the
AdS action (\ref{(A)dS2n+1}) is supplemented with a boundary term of the form
\begin{equation}
B_{2n}=-\kappa n\int\limits_0^1dt\int\limits_0^t ds\;\epsilon \theta e\left( \widetilde{R}+t^2 \theta^2+s^2 e^2\right)^{n-1},  \label{B2n}
\end{equation}
where $\widetilde{R}$ and $\theta$ are the intrinsic and extrinsic curvatures of the boundary. The resulting action attains an extremum for boundary
conditions that fix the extrinsic curvature of the boundary. In that reference it is also shown that this action principle yields finite charges (mass, angular momentum) without resorting to ad-hoc regularizations or background subtractions. It can be asserted that in this case --as in many others--, the demand of gauge invariance is sufficient to cure other seemingly unrelated problems. 

\subsubsection{Transgressions}    

The boundary term (\ref{B2n}) that that ensures convergence of the action and charges turns out to have other remarkable properties. It makes the action gauge invariant --and not just quasi-invariant-- under gauge transformations that keep both, the intrinsic AdS geometry, and the extrinsic curvature, fixed at the boundary. The condition of having a fixed asymptotic AdS geometry is natural for localized matter distributions such as black holes. Fixing the extrinsic curvature, on the other hand, implies that the connection approaches a fixed reference connection at infinity in a prescribed manner. 

On closer examination, this boundary term can be seen to convert the action into a \textit{transgression}, a mathematically well-defined object. A transgression form is a gauge invariant expression whose exterior derivative yields the difference of two Chern classes \cite{Nakahara},
\begin{equation}
 d{\cal T}_{2n-1} (A, \bar{A})={\cal P}_{2n}(A)-{\cal P}_{2n}(\bar{A}),  \label{transgression}
\end{equation}
where $A$ and $\bar{A}$ are two connections in the same Lie algebra. There is an explicit expression for the transgression form in terms of the Chern-Simons forms for $A$ and $\bar{A}$, 
\begin{equation}
{\cal T}_{2n+1} (A, \bar{A})={\cal C}_{2n+1}(A)-{\cal C}_{2n+1}(\bar{A}) + d{\cal B}_{2n}(A,\bar{A}). \label{transg}
\end{equation}
The last term in the R.H.S. is uniquely determined by the condition that the transgression form be invariant under simultaneous gauge transformations of both connections throughout the entire manifold $M$
\begin{eqnarray}
A \rightarrow A'=\Lambda^{-1} A \Lambda + \Lambda^{-1}d\Lambda \\ \label{delta A}
\bar{A} \rightarrow \bar{A}'=\bar{\Lambda}^{-1} \bar{A} \bar{\Lambda} + \bar{\Lambda}^{-1}d\bar{\Lambda} \label{delta barA}
\end{eqnarray}
with the matching condition at the boundary,
\begin{equation}
\bar{\Lambda}(x)-\Lambda(x) = 0, \;\; \mbox{ for }\;\;  x \in \partial M. \label{matching}
\end{equation}
It can be seen that the boundary term in (\ref{B2n}) is precisely the boundary term ${\cal B}_{2n}$ in the transgression form. The interpretation now presents some subtleties. Clearly one is not inclined to duplicate the fields by introducing a second dynamically independent set of fields ($\bar{A}$), having exactly the same couplings, gauge symmetry and quantum numbers. 

One possible interpretation is to view the second connection as a nondynamical reference field. This goes against the principle according to which every quantity that occurs in the action that is not a coupling constant,  mass parameter, or numerical coefficient like the dimension or a combinatorial  factor, should be a dynamical quantum variable \cite{Yang}. Even if one accepts the existence of this unwelcome guest, an explanation would be needed to justify its not being seen in nature. However, other possibilities exists, as we discuss next.

\subsubsection{Cobordism}   

An alternative interpretation could be to assume that the spacetime is duplicated and we happen to live in one of the two parallel words where $A$ is present, while $\bar{A}$ lives in the other. An obvious drawback of this interpretation is that the action for $\bar{A}$ has the wrong sign, which would lead to ghosts or rather unphysical negative energy states. We could ignore this fact because the ghosts would live in the ``parallel universe'' that we don't see, but this is not completely true because, at least at the boundary, the two universes meet.

Interestingly, $A$ and $\bar{A}$ only couple through the boundary term, ${\cal B}_{2n}(A,\bar{A})$, and therefore, the bulk where $A$ is defined need not be the same one where $\bar{A}$ lives. These two worlds must only share the same boundary, where condition (\ref{matching}) makes sense; they are independent but \textit{cobordant} manifolds.

The negative sign in front of ${\cal C}(\bar{A})$ is an indication that the orientation of the parallel universe must be reversed. Then, the action can be written as
\begin{equation}
 I[A,\bar{A}]= \int_{M} {\cal C}(A) +\int_{\bar{M}} {\cal C}(\bar{A}) +\int_{\partial M} {\cal B}(A, \bar{A}),
\end{equation}
where the orientation of $\bar{M}$ has been reversed so that at the common boundary, $\partial M=-\partial \bar{M}$. In other words, the two cobordant manifolds $M$ and $\bar{M}$, with the new orientation, define a single, uniformly oriented surface sewn at the common boundary which is also correctly oriented.

The picture that emerges in this interpretation is one where we live in a region of spacetime ($M$) characterized by the dynamical field $A$. At he boundary of our region, $\partial M$, there exists another field with identical properties as $A$ and matching gauge symmetry. This second field $\bar{A}$ extends on to a cobordant manifold $M$, to which we have no direct access except through the interaction of $\bar{A}$ with our $A$\cite{MOTZ2}. If the spacetime we live in is asymptotically AdS, this could be a reasonable scenario since the boundary then is causally connected to the bulk and can be easily viewed as the common boundary of two --or more-- asymptotically AdS spacetimes \cite{Zanelli2006}.

\section{CS as brane couplings}    

Although CS forms appeared in high energy physics more than 30 years ago, in order to achieve local supersymmetry \cite{CJS} and as Lagrangians in quantum field theory \cite{Schonfeld,Deser-Jackiw-Templeton}, recently a different use has been identified. As noted in Sect. 1, under gauge transformation CS forms change essentially like an abelian connection. Hence, they can be used like the electromagnetic vector potential $A$, to couple to a conserved current. However, since the CS forms have support on a ($2n+1$)-dimensional manifold, they couple gauge fields to extended sources ($2n$-dimensional membranes), charged with respect to some gauge color \cite{z,ez,mz}. 

The CS form provides a gauge-invariant coupling between the gauge potential and an extended source in a consistent manner, something that could be achieved by the ``more natural" minimal couplings $A_{\mu_1 \mu_2 \cdots \mu_p} j^{\mu_1 \mu_2 \cdots \mu_p}$, only for an abelian potential \cite{Teitelboim:1985ya}.

\subsection{Minimal coupling and geometry} 

The other common situation in which total derivatives play a fundamental role is in the coupling between a field and a conserved current.\footnote{The following discussion is based on \cite{EAV2011}.}  The epitome of such coupling is the interaction Lagrangian between the electromagnetic field and an electric current,
\begin{equation}
I_{\mbox{\tiny{Int}}}=\int_{\Gamma} A_{\mu}j^{\mu}d^4x\, ,
\label{minicoupl}
\end{equation} 
where $A_{\mu}$ is the vector potential and $j^{\mu}$ is the electromagnetic 4-vector current density. This coupling has two nontrivial properties besides its obvious Lorentz invariance, gauge invariance, and metric independence, both of which are common to all CS forms.

As mentioned in Sect.1, invariance of $I_{\mbox{\tiny{Int}}}$ under gauge transformations is a consequence of two properties of the current: its gauge invariance, $j'^{\mu}=j^{\mu}$ --as physical observables should--, and its conservation, $\partial_{\mu}j^{\mu}=0$. Conversely, the minimal coupling means that the vector potential \textit{can only couple consistently} (in a gauge invariant way) to a gauge-invariant, conserved current. This statement makes no reference to the equations of motion of the charges or to Maxwell's equations. Additionally, gauge invariance of the action implies the conservation of electric charge. The conserved charge is precisely the generator of the symmetry that implies its conservation. 

The metric independence of the minimal coupling has a more subtle meaning. It can be trivially verified from the fact that under a change of coordinates, $A_{\mu}$ transforms as a covariant vector while $j^{\mu}$ is a vector density and, therefore, the integrand in (\ref{minicoupl}) is coordinate invariant.  No ``$\sqrt{|g|}$" is required in the integration measure, which implies that the same coupling can be used in a curved background or in flat space and in any coordinate frame. This ultimately means that the integrand in (\ref{minicoupl}) is an intrinsic property of the field $A$ over the world line swept by the point charges in their evolution.

In the case of one point charge, the current is best understood as the dual the three-form delta function (a density) supported on the worldline of the particle, $j=e*\delta(\Gamma)$ where $e$ is the electric charge. Hence, $A_{\mu}j^{\mu}d^4x=eA\wedge \delta(\Gamma)$, and (\ref{minicoupl}) reads
\begin{equation}
I_{\mbox{\tiny{Int}}}=e\int_{\Gamma} A_{\mu}(z)dz^{\mu}=e\int_{\Gamma} A\, ,
\label{minicoupl2}
\end{equation} 
where the coordinate $z^{\mu}$ is any convenient parametrization of the worldline $\Gamma$. This is correct since distributions are elements in the dual of a space of test functions, that upon integration yield numbers; in the case of the Dirac delta, it yields the value of the test function at the support, which is exactly the content of the equivalence between (\ref{minicoupl}) and (\ref{minicoupl2}). In this case, the current merely projects the one-form $A$ defined everywhere in spacetime, onto the worldline. The result is clearly independent of the metric of the ambient spacetime and of the metric of the worldline, which in this simple case corresponds to the choice of coordinate $z^{\mu}$. 

It is reassuring that not only the coupling, but also the conservation law $\partial_{\mu}j^{\mu}=0$ doesn't require a metric,  since $\partial$ is the ordinary derivative, and $j$ is a contravariant vector density, which makes the conservation equation valid in any coordinate basis and for any metric.  Metric independence ultimately means that the coupling is insensitive to deformations of the worldline of the charge, and of the spacetime metric. Thus, regardless of how the particle twists and turns in it evolution, or the metric properties of spacetime where the interaction takes place, the coupling remains consistently gauge invariant. This fact is crucial for the dynamical consistency of the coupling to membranes or other extended objects. 

\subsection{Extended sources and CS couplings}  

A ($2p+1$)-CS form describes the coupling between a connection $\mathbf{A}$ and a membrane whose time evolution sweeps a ($2p+1$)-dimensional volume. The consistency of this scheme follows from the precise form of the coupling \cite{Edelstein:2010sh},  
\begin{equation}
I[\mathbf{A}; \mathbf{j}]=\int \langle \mathbf{j}_{2p} \wedge \mathfrak{C}_{2p+1}
(\mathbf{A})\rangle,  \label{general CS coupling}
\end{equation}
where $\mathfrak{C}_{2p+1}$ is the algebra-valued form whose trace is the ($2p+1$)-CS form living on the brane history, $\mathcal{C}_{2p+1} (\mathbf{A})= \langle \mathfrak{C}_{2p+1} (\mathbf{A})\rangle$. The current generated by the $2p$-brane is represented by the ($D-2p-1$)-form $\mathbf{j}$ supported on the worldvolume of the brane,
\begin{equation}
\mathbf{j}_{2p}=qj^{a_1 a_2 \cdots a_s} \mathbf{K}_{a_1} \mathbf{K}_{a_2}\cdots \mathbf{K}_{a_s} \delta(\Gamma) dx^{\alpha_1}\wedge dx^{\alpha_2}\wedge \cdots dx^{\alpha_{D-2p-1}}, \label{j}
\end{equation}
where $dx^{\alpha_i}$ are transverse directions to $\Gamma$. The integration over the $D-2p-1$ transverse directions yields a ($2p+1$)-CS form integrated over the worldvolume of the $2p$-brane. The invariance under the gauge transformations can be checked directly by noting that, $I[\mathbf{A}; \mathbf{j}]$ changes by a (locally) exact form, provided the current is covariantly conserved,
\begin{equation}
D \mathbf{j}_{2p}= d\mathbf{j}_{2p} + [\mathbf{A}, \mathbf{j}_{2p}] =0, \label{dj}
\end{equation}
which can be independently checked for (\ref{j}). Moreover, if the current $\mathbf{j}_{2p}$ results from particles or fields whose dynamics is governed by an action invariant under the same gauge group $\mathbf{G}$, then its conservation is guaranteed by consistency. On the worldvolume, however, the gauge symmetry is reduced to the subgroup that commutes with $j^{a_1 a_2 \cdots a_s} \mathbf{K}_{a_1} \mathbf{K}_{a_2}\cdots \mathbf{K}_{a_s}$. 

An interesting --and possibly the simplest-- example of such embedded brane occurs when an identification is made in the spatial slice of AdS$_3$, using a rotational Killing vector with a fixed point. In that case, a deficit angle is produced and the conical geometry produced around the singularity can be identified with a point particle \cite{Deser-et-al}; the  singularity is the worldline of the particle where the curvature behaves as a Dirac delta. The geometry is analogous to that of the BTZ black hole \cite{Banados:1992wn}, but the naked singularity results from a wrong sign in the mass parameter of the solution \cite{Miskovic:2009uz}. A similar situation arises also when one considers a co-dimension 2 brane in higher dimensions \cite{mz}. In all these cases it is confirmed that the coupling between this 0-brane and the (nonabelian) connection is indeed of the form (\ref{general CS coupling}). These branes only affect the topological structure of the geometry, but the local geometry outside the worldline of the source remains unchanged.

\subsection{3D-CS systems and condensed matter}  

There may be more realistic situations where a three-dimensional CS theory could give rise to interesting  effects. For example, materials in ordinary four-dimensional spacetime whose excitations propagate on two-dimensional layers, display a quantum Hall effect responsible for superconductivity of high critical temperature \cite{Zhang-et-al,Balachandran}. 

A (2+1)-dimensional CS system is naturally generated at the interface separating two regions of three-dimensional space in which a Yang-Mills theory has different vacua. Consider an action of the form
\begin{equation}
I[A]= \frac{1}{2} \int_M \left( Tr[F\wedge *F] -  \Theta(x) Tr[F\wedge F] \right), \label{theta-action}
\end{equation}
where the last term has the form of a topological invariant, but it fails to be topological precisely because it is multiplied by a function. If this function $\Theta$ takes a constant value $\theta_1$ in the region $\tilde{M} \subset M$ and $\theta_2$ elsewhere, then the second term can also be written as a coupling between the Chern-Simons and a surface current,
\begin{equation}
\int_M \frac{\Theta }{2}Tr[F\wedge F]=\int_{\partial\tilde{M}} jTr[\wedge A\wedge dA]  ,  \label{theta-CS}
\end{equation}
where the surface current is the one-form $\ j=d \Theta=(\theta_1 - \theta_2) \delta (\Sigma)dz$, and $z$ is the coordinate along outward normal to the surface of $\tilde{M}$, $\Sigma =\partial \tilde{M}$. Since the $\theta$-term is locally exact, the field equations, both inside and outside $\tilde{M}$ are the same as in vacuum. However, this term modifies the behavior of the field \textit{at} the surface $\Sigma $.

This coupling between the CS form and the spacetime boundary of the region $\tilde{M}$ has physical consequences even in the simple case of the abelian theory (electromagnetism). Although Maxwell's equations are not affected by the $\theta$-term, the interface rotates the polarization plane of an electromagnetic wave that crosses it \cite{Huerta-z}. Another effect of such coupling is a modification of the Casimir energy inside an empty region surrounded by a ``material" in which $\theta\neq 0$ \cite{Canfora-R-z}. It turns out that the so-called topological insulators are materials that produce the effect of modifying the ``$\theta$ vacuum" of electromagnetic theory, a phenomenon that is attracting considerable attention of the condensed matter community \cite{Vozmediano}.

\section{A cursory look at supergravity}    

The scope of this review is gravity as a gauge theory for the local Lorentz symmetry and its ``natural" extensions, the Poincar\'e, de Sitter, and anti-de Sitter groups. Rigid (i.e., not gauged) supersymmetry (\textbf{SUSY}) was identified as the only nontrivial way to combine the Poincar\'e symmetry of flat spacetime and the internal symmetries, reflected by the conservation laws (selection rules)
of elementary particles, that is not simply a direct product of groups. 

Most theoretical physicists view SUSY as a legitimate --even desirable-- feature of nature. The reason for its popularity rests on its uniqueness and elegance, mixing bosons (integer spin particles) and fermions (half integer spin particles) as different aspects of a unified system. SUSY theories have better renormalizability properties than non-SUSY ones and prevents coupling constants from getting renormalized, two welcome features in the elementary particle models. Sadly, however, SUSY has never been experimentally observed, not even as a rough approximation.

The most intriguing --and uncomfortable-- aspect of SUSY is that it predicts the existence of fermionic carriers of interactions and bosonic constituents of matter. The fact that such particles have never been observed and blatant the evidence that bosons and fermions play such radically different roles in nature, strongly indicates that SUSY must be badly broken at the scale of our observations. On the other hand, there is no clear mechanism at present to break SUSY without losing its benefits.

The improved renormalizability of SUSY is particularly attractive in the context of gravity: the supersymmetric extension of GR (\textbf{SUGRA}) whose ultraviolet divergences exactly cancel at the one-loop level \cite{SUGRA, PvN}. There is yet another beautiful feature of SUSY: local (gauge) SUSY is not only compatible with gravity but, by consistency, it \textit{requires} gravity \cite{Sohnius, Freund}.

\subsection{Supergravity as a gauged SUSY}   

From an algebraic point of view, SUSY is a graded Lie algebra, having both commutators ($\left[\cdot, \cdot\right ]$) and anticommutators ($\left\{ \cdot ,\cdot \right\} $), also called a superalgebra. See, e. g., \cite{Sohnius, Freund} for a review of this vast topic. In physics, SUSY appears naturally as a only nontrivial extension of the translation group, where the successive action of two supersymmetry transformations is defined as
\begin{equation}
[\bar{Q}\epsilon, \bar{Q}\eta] \sim (\bar{\epsilon} \gamma^\mu \eta -\bar{\eta}\gamma^\mu \epsilon) \partial_\mu \,.
\label{susy-0}
\end{equation}
This identifies supersymmetry as ``the square root of the translation" group. If one additionally demands that $Q$ be in a spinor representation, the supersymmetric algebra becomes a natural extension  of the Poincar\'e group. The underpinnings of SUSY can be found in the possibility of extending a given Lie algebra $\mathcal{G}_0$ into a graded algebra, by the inclusion of fermionic generators which complete a graded algebra $\mathcal{G}_0 \supseteq \mathcal{G}_0$.There is nothing about the Poincar\'e algebra that links it uniquely to supersymmetry, although this is historically the first area of its application. Graded Lie algebras can be constructed for all classical groups most often used in physics.

The fact that the supersymmetry generators close on the translation group makes it immediately natural to assume that the local version of (\ref{susy-0}) in which $\epsilon$ and $\eta$ are fields, implies local translations appear on the right hand side. Loosely speaking, local SUSY requires a gauge theory for the translation group. As local translations can be identified with coordinate diffeomorphisms, a theory invariant under gauge SUSY was naturally identified as containing gravity: \textit{supergravity}. The dominant view over the past four decades has been that gauge supersymmetry can only be supergravity, where gravity is understood as a theory obtained by gauging the translation group, a superficially very plausible idea. 

On closer scrutiny, however, this argument has a number of holes. First, the invariance under diffeomorphisms is not an exclusive feature of gravity: any action, for any local field theory, is invariant under diffeomorphisms. Second, the definition of spinors and Clifford algebras is related to representations of orthogonal groups $SO(n,m)$, and therefore has no direct connection with spacetime, except for very special cases like maximally symmetric manifolds (Minkowski, (a)dS). These spaces are extremely exceptional solutions of gravitation theory, certainly not realized in general relativity and much less in quantum gravity. 

A more technical point is that the generators of diffeomorphisms, the Hamiltonian constraints $\mathcal{H}_{\perp}$ and $\mathcal{H}_i$ form an \textit{open algebra}, which is analogous to a Lie algebra but in which the structure constants are replaced by fields, and in the case of gravity, by dynamical fields. This makes the symmetry dependent on the states, something rather unmanageable except for very few applications, but certainly unlikely to yield a consistent quantum picture. Finally, most supergravity theories require using some field equations in order to close the SUSY algebra, but this again is hard to assume consistently in a quantum version of the theory, where the classical equations are irrelevant.

\subsection{Supergravity as a CS theory for a superalgebra}   
As we have argued above, the CS construction only requires a Lie algebra and an odd-dimensional manifold. It turns out that in any dimension, the supersymmetric extensions of the anti-de Sitter group $SO(D,2)$ are completely determined and therefore, the CS theories for those superalgebras are uniquely fixed. In odd dimensions, the corresponding superalgebras are given in the following table\footnote{The algebra $osp(p|q)$ (resp. $usp(p|q)$) is that which generates the orthosymplectic (resp. unitary-symplectic) Lie group. This group is defined as the one that leaves invariant the quadratic form $G_{AB}z^A z^B = g_{ab}x^a x^b + \gamma_{\alpha \beta} \theta^{\alpha} \theta^{\beta}$, where $g_{ab}$ is a $p$-dimensional symmetric (resp. hermitean) matrix and $\gamma_{\alpha \beta}$ is a $q$-dimensional antisymmetric (resp. anti-hermitean) matrix.}  \cite{vH-VP, TrZ1}

\begin{center}
\vspace{0.2cm} 
\begin{tabular}{|l|c|c|}
\hline
D & Super Algebra & Conjugation Matrix \\ \hline
3 mod 8 & $osp(m|N)$ & $C^{T}=-C$ \\ \hline
7 mod 8 & $osp(N|m)$ & $C^{T}=C$ \\ \hline
5 mod 4 & $usp(m|N)$ & $C^{\dag }=C$ \\ \hline
\end{tabular}
\end{center}
\textbf{Table 3:} Superalgebras for different (odd) dimensions and the corresponding conjugation matrices.\\
\
Here $m=2^{[D/2]}$ is the dimension of the spinor representation in dimension $D$. The different algebras depend on the properties of the Clifford algebra of the corresponding dimension and the symmetry properties the charge conjugation matrix $C_{ab}$, as indicated in the third column. The construction of these algebras is elementary and can be found in \cite{TrZ2}.

The resulting CS actions include the CS action for gravity in the appropriate dimension, as discussed in sections  4.3. In addition, there are some nonabelian gauge fields also described by a CS action, coupled to the fermionic fields. The important fact is that the Lagrangian always defines a gauge theory invariant under the corresponding local transformations. The SUSY transformations are of the form
\begin{center}
$\begin{array}{ll}
\delta e^a=\frac{1}{2}\bar{\epsilon}^i\Gamma^a \psi_i, & \delta \omega^{ab}= \frac{-1}{2} \bar{\epsilon}^i\Gamma^{ab}\psi_i \\
\delta \psi_i=D\epsilon_i, & \delta a^i_{\; j}= \bar{\epsilon}^i \psi_j - \bar{\psi}^i \epsilon_j ,
\end{array}$
\end{center}
where the indices $1\leq i,j \leq \mathcal{N}$ correspond to a vector representation of the local internal R-symmetry, whose connection is represented here by $ a^i_{\; j}$.
The resulting CS supergravities are in general different from the standard versions of supergravity obtained by the supersymmetric extension of the Einstein-Hilbert Lagrangian. The following table compares the field contents of these two forms of extended supergravities for some dimensions,
\begin{center}
\vspace{0.2cm} 
\begin{tabular}{|l|l|l|c|}
\hline
$D$ & Standard SUGRA & CS SUGRA & Algebra \\ 
\hline
5 & $e_{\mu}^a \;\psi_{\mu}^{\alpha}\;\bar{\psi}_{\alpha \mu}$ & $e_{\mu}^a\;\omega_{\mu}^{ab}\;A_{\mu}\;a_{\mu j}^i\; \psi_{i\mu}^{\alpha} \;\bar{\psi}_{\alpha \mu}^i $ & $usp(2,2|\mathcal{N})$ \\ 
\hline
7 & $e_{\mu}^a\;A_{[3]}\;a_{\mu j}^i\;\lambda^{\alpha}\;\phi\;\psi_{\mu}^{\alpha i}$ & $e_{\mu}^a\;\omega_{\mu}^{ab}\;a_{\mu j}^i
\;\psi_{\mu}^{\alpha i}\, $ & $osp(\mathcal{N}|8) \, (\mathcal{N}=2n)$ \\ 
\hline
11 & $e_{\mu}^a \;A_{[3]}\; \psi_{\mu}^{\alpha} \,$ & $e_{\mu}^a\;\omega_{\mu}^{ab}\; b_{\mu}^{abcde}\;\psi_{\mu}^{\alpha}\;$ & $osp(32|\mathcal{N})$ \\
\hline
\end{tabular}
\end{center} 
\textbf{Table4:} Comparison between the field contents of standard and CS supergravities for different odd dimensions.\\
\
A generic feature of all these supergravities is that they contain spin-one gauge and spin-two bosonic and spin-three half fermionic fields, and no higher spin fields. This could be regarded as a virtue in view of the difficulty to define consistent couplings of fields of spin larger than two. However, the CS ``trick" seems to allow for higher spins as well, and might be the underlying structure of the recent developments in higher spin field theories \cite{Engquist-Hohm}. 

For completeness one should also mention that the CS construction can also be extended to some cases where the spin 3/2 field is replaced by a combination of a spin 1/2 and the vielbein, producing a locally supersymmetric theory without gravitini \cite{AVZ}.

\subsection{Relation between CS and standard SUGRAs}   
One obvious question that one can ask is whether the CS supergravities can be related to the standard SUGRAs: is one a particular sector or a  truncation of the other? An indication of this possibility comes from the fact that the Poincar\'e algebra, that presumably underlies the nearly flat limit of AdS gravity, could be obtained in the $\Lambda\rightarrow 0$ limit of the AdS algebra. This In\"on\"u-Wigner contraction \cite{Inonu} is a particular case of a more general method to obtain non-semisimple Lie algebras from simple or semisimple ones, that go by the names of contractions and expansions first discussed in a physical context in \cite{Hatsuda-etal} and further developed in \cite{deAzcarraga-etal1}(see, \cite{deAzcarraga-rev} for a detailed review of the method). 

The general pattern to relate the Lagrangian of a standard SUGRA in $D$ dimensions to the CS supergravity for the super-AdS$_D$ algebra is far from straightforward \cite{EHTZ}. A remarkable proof of the nontrivial character of this relation is provided by the recent elucidation in \cite{Izaurieta-Rodriguez} of the nontrivial truncation that is needed to obtain the standard 11-dimensional supergravity of Cremmer, Julia and Scherk \cite{CJS} from the CS Lagrangian for osp(1|32), the simplest CS supergravity in 11-dimensions of \cite{TrZ1}. The expansion method was further exploited in \cite{deAzcarraga-etal2} to relate different CS supergravities in 2+1 dimensions, where the the nontrivial nature of those relations, even in this seemingly simpler setting, is explicitly shown.

\section{Summary}    

The relevance of gauge symmetry in physics cannot be overemphasized. One of the great achievements of physics in the last century was to establish that all interactions in nature are based on gauge invariance. The fact that nature possesses this fundamental symmetry explains the binding forces in the atomic nucleus, the functioning of stars, the chemistry that supports life, and the geometry of the universe. This unifying principle is comparable to the invention of mechanics in the XVII century, or electrodynamics and statistical mechanics of the XIX century. 

It is a remarkable feature rooted in the equivalence principle, that gravity is a gauge theory for the Lorentz group. The spacetime geometry can be described by two independent notions, metricity and affinity, each one described by a fundamental field that transforms under the local symmetry in a definite representation. It is an even more remarkable feature, that in odd dimensions these two fundamental objects can combine to become a connection for an enlarged gauge symmetry, the (anti-) de Sitter or the Poincar\'e groups. The resulting theory is described by a CS form that has no arbitrary free parameters, no dimensionful couplings, and whose gauge invariance is independent of the spacetime geometry. 

None of this seems random. One cannot help feeling that something profound and beautiful lies in these structures. Whether the CS theories of gravity, or their more ambitious supersymmetric extensions turn out to be the way to understand the connection between gravitation and quantum mechanics, remains to be seen. However, the fact that CS forms are singled out in gravity, the fact that they play such an important role in the couplings between gauge fields and sources, their deep relation with quantum mechanics, strongly suggests that there is some meaning to it. This doesn't look like a contingent result of natural chaos.

Ivar Ekeland reflects on the sense of nature as is revealed to us: \textit{But is contingency complete or is there room for meaning? Must we be content to merely note the facts, or should we look for reasons? De events follow one another randomly, or does the world function according to certain rules that we can reveal and make use of? We often don't like the way things are, and some people go so far as to give their lives to change them; the quest for meaning must therefore be part of human existence} \cite{Ekeland}.

We may not go as far as to give our lives to convince anyone of the virtues of Chern-Simons theories. But CS forms surely make our quest for meaning a more aesthetic endeavour.

--------------------- \\

\textbf{Acknowledgments:}
Many enlightening discussions with Pedro Alvarez, Eloy Ay\'on-Beato, Aiyalam P. Balachandran, Claudio Bunster, Fabrizio Canfora, Gast\'on Giribet, Jos\'e Edelstein, Luis Huerta, Olivera Mi\v{s}kovi\'c, Ricardo Troncoso, and Mauricio Valenzuela, are warmly acknowledged. This work has been partially supported by FONDECYT grants 1100755, 1085322, 1100328, 1110102, and by the Southern Theoretical Physics Laboratory ACT- 91 grants from CONICYT. The Centro de Estudios Cient\'{\i}ficos (CECs) is funded by the Chilean Government through the Centers of Excellence Base Financing Program of CONICYT. 

\section*{Appendix A. First order fields}  

\subsection*{A.1   The vielbein}  

Spacetime is a smooth $D$-dimensional manifold $M$, of Lorentzian signature $(-1,1,1,\cdots,1)$. At every point on $x \in M$ there is a $D$-dimensional tangent space $T_{x}$, which is a good approximation of the manifold $M$ in the neighbourhood of $x$. This tangent space is the reference frame of a freely falling observer mentioned in the Equivalence Principle. The fact that the measurements carried out in any reference frame in spacetime can be translated to those in a freely falling frame, means that  there is an isomorphism between tensors on $M$ by tensors on $T_{x}$, represented by means of a linear mapping, also called {\it ``soldering form''} or {\it ``vielbein''}. It is sufficient to define this mapping on a complete set of vectors such as the coordinate separation $dx^{\mu }$ between two infinitesimally close points on $M$. The corresponding separation in $T_{x}$ is defined to be
\begin{equation}
dz^{a}=e_{\mu }^{a}(x)dx^{\mu } , \label{vielbein}
\end{equation}
where $z^a$ represent an orthonormal coordinate basis in the tangent space. For this reason the vielbein is also viewed as a local orthonormal frame. Since $T_x$ is a standard Minkowski space, it has a natural metric, $\eta_{ab}$, which defines a metric on $M$ through the isomorphism $e^a_{\mu}$. In fact,
\begin{eqnarray}
\nonumber
ds^2&=&\eta_{ab} dz^a dz^b \\ \nonumber
&=&\eta_{ab}\, e_{\mu}^a(x) dx^{\mu}\, e_{\nu}^b(x) dx^{\nu} \\
&=&g_{\mu \nu}(x)dx^{\mu} dx^{\nu},
\end{eqnarray}
where 
\begin{equation}
\eta_{ab}\, e_{\mu}^a(x)\, e_{\nu}^b(x)\equiv g_{\mu \nu}(x)\, , 
\label{metric}
\end{equation}
is the metric on $M$, induced by the vielbein $e_{\mu}^a(x)$ and the tangent space metric $\eta_{ab}$.

This relation can be read as to mean that the vielbein is in this sense the square root of the metric. Given $e_{\mu }^{a}(x)$ one can find the metric and therefore, all the metric properties of spacetime are contained in the vielbein. The converse, however, is not true: given the metric, there exist infinitely many choices of vielbein that reproduce the same metric. 

The definition (\ref{vielbein}) makes sense only if the vielbein $e_{\mu }^{a}(x)$ transforms as a covariant vector under diffeomorphisms on $M $ and as a contravariant vector under local Lorentz rotations of $T_{x}$, $SO(D-1,1)$, as
\begin{equation}
e_{\mu }^{a}(x)\longrightarrow e_{\mu}^{\prime a}(x)=\Lambda_{b}^{a}(x)e_{\mu}^{b}(x),  \label{Transf-e}
\end{equation}
where the matrix ${\bf \Lambda}(x)$ leaves the metric in the tangent space unchanged,
\begin{equation}
\Lambda_{c}^{a}(x)\Lambda_{d}^{b}(x)\eta_{ab}=\eta_{cd}, \label{Lorentz}
\end{equation}
then the metric $g_{\mu \nu }(x)$ is clearly unchanged. The matrices that satisfy (\ref{Lorentz}) form the Lorentz group $SO(D-1,1)$. This means, in particular, that there are many more components in $e_{\mu }^{a}$ than in $g_{\mu \nu }$. In fact, the vielbein has $D^{2}$ independent components, whereas the metric has only $D(D+1)/2$. The mismatch is exactly $D(D-1)/2$, the number of independent rotations in $D$ dimensions.

Every tangent space at one point of the manifold is a Minkowski space identical to all others, each one invariant under the action of the Lorentz group. This endows the spacetime manifold with a fibre bundle structure, the \textit{tangent bundle}, where the basis is the spacetime and the fibres are the tangent spaces on which the Lorentz group acts locally. This can also be regarded as a \textit{collection of vector spaces  parametrized by the manifold}, $\{T_x, x\in M\}$. Either way, the essential point is that the manifold $M$, labelled by the coordinates $x^\mu$ is the spacetime where we live, and the collection of tangent spaces over it is where the symmetry group acts.

\subsection*{A.2   The Lorentz Connection\footnote{In physics, this is often called the \textit{spin connection}. The word ``spin" is due to the fact that it arises naturally in the discussion of spinors, which carry a special representation of the group of rotations in the tangent space, but that is irrelevant here. For a more extended discussion, there are several texts such as those of Refs.\cite{Nakahara}, \cite{Schutz}, \cite{Eguchi-Gilkey-Hanson} and \cite{Goeckeler-Schuecker}} }  

In order to define a derivative on the manifold, a connection is required so that the differential structure remains invariant under local Lorentz transformations $\Lambda(x)$, even if they act independently at each spacetime point. This is achieved by introducing the Lorentz connection, $\omega_{\;b\mu}^{a}$, which is defined so that if $\phi^{a}(x)$ is a field in a vector representation of the Lorentz group, its covariant derivative,
\begin{equation}
D_{\mu}\phi^{a}(x)=\partial _{\mu }\phi^{a}(x)+\omega _{\;b\mu}^{a}(x) \phi^{b}(x),  \label{D-mu}
\end{equation}
also transforms like a Lorentz vector at $x$, provided $\omega$  transforms as a connection,
\begin{equation}
\omega _{\;b\mu }^{a}(x)\longrightarrow \omega_{\;b\mu }^{\prime a}(x)= \Lambda_{c}^{a}(x)\Lambda _{b}^{d}(x)\omega_{\;d\mu }^{c}(x)+\Lambda_{c}^{a}(x)\partial _{\mu }\Lambda_{b}^{c}(x). \label{LorConn}
\end{equation}

The connection $\omega_{\;b\mu }^a(x)$ defines the \textit{parallel transport} of Lorentz tensors between tangent spaces at nearby points, $T_{x} \rightarrow T_{x+dx}$. If $\phi^{a}(x)$ is a  vector field, its parallel-transported from $x+dx$ to $x$, is defined as
\begin{eqnarray}
\phi_{||}^{a}(x) &\equiv& \phi^{a}(x+dx)+ dx^{\mu}\omega_{\;b\mu}^{a}(x)\phi^{b}(x) \\ \nonumber &=& \phi^{a}(x) + dx^{\mu}[\partial_{\mu} \phi^{a}(x)+ \omega_{\;b\mu}^{a}(x)\phi^{b}(x)] .
\label{ParallelTransp}
\end{eqnarray}
Consequently, the covariant derivative measures the change in a tensor produced by parallel transport between neighbouring points,
\begin{eqnarray}
dx^{\mu }D_{\mu }\phi^{a}(x) &=& \phi_{||}^{a}(x)-\phi^{a}(x) \\ \nonumber &=&dx^{\mu}[\partial_{\mu} \phi^{a}+\omega_{\;b\mu }^{a}(x)\phi^{b}(x)]
\label{Lor-cov-der}.
\end{eqnarray}
This notion of parallelism is analogous to the one defined for vectors whose components are referred to a coordinate basis,
\begin{equation}
\varphi_{||}^{\mu}(x) = \varphi^{\mu}(x) + dx^{\lambda}[\partial_{\lambda} \varphi^{\mu}(x)+ \Gamma_{\lambda \rho}^{\mu}(x) \varphi^{\rho}(x)] .
\end{equation}
These two definitions are independent as they refer to objects on different spaces, but they could be related using the soldering $e^a_{\mu}$, between the base manifold and the tangent space.

The group $SO(D-1,1)$ has two invariant tensors, the Minkowski metric, $\eta_{ab}$, and the totally antisymmetric Levi-Civitta tensor, $\epsilon_{a_1 a_2 \cdots a_D}$. Because they are the same in every tangent space, they are constant ($\partial_\mu \eta_{ab}=0$, $\partial_\mu \epsilon_{a_1 a_2 \cdots a_D}=0$) and since they are also invariant, they are covariantly constant,
\begin{eqnarray}
d\eta_{ab}=D\eta_{ab}&=&0, \\
d\epsilon_{a_1 a_2 \cdots s_D}=D\epsilon_{a_1 a_2 \cdots a_D} &=&0.
\end{eqnarray}
This implies that the Lorentz connection satisfies two identities,
\begin{eqnarray}
\eta_{ac} \omega^c_{\;\;b} &=& -\eta_{bc} \omega^c_{\;\;a}, \label{metricity} \\
\epsilon_{b_1 a_2 \cdots a_D}\omega^{b_1}_{\;\;a_1}+\epsilon_{a_1 b_2 \cdots a_D}\omega^{b_2}_{\;\;a_2}+ &\cdots& + \epsilon_{a_1 a_2 \cdots b_D}\omega^{b_D}_{\;\;a_D} =0 \label{Chebichev}.
\end{eqnarray}
The requirement that the Lorentz connection be compatible with the metric structure of the tangent space (\ref{metricity}) restricts $\omega^{ab}$ to be antisymmetric, while the second relation (\ref{Chebichev}) does not impose further restrictions on the components of the Lorentz connection. Then, the number of independent components of $\omega _{\;b\mu }^{a}$ is $D^{2}(D-1)/2$, which is less than the number of independent components of the Christoffel symbol ($D^{2}(D+1)/2$).

\subsection*{A.3   Curvature}

The 1-form exterior derivative operator, $dx^{\mu}\partial_{\mu}\wedge$ is such that acting on a p-form, $\alpha_p$, it yields a (p+1)-form, $d\alpha_p$. One of the fundamental properties of exterior calculus is that the second exterior derivative of a differential form vanishes identically,
\begin{equation}\label{d2}
d(d\alpha_p) =: d^2 \alpha_p=0.
\end{equation}
This is trivially so on continuously differentiable forms. A consequence of this is that the square of the covariant derivative operator is not a differential operator, but an algebraic operator, the curvature two-form. For instance, the second covariant derivative of a vector yields
\begin{eqnarray}\label{D2}
D^2\phi^a &=& D[d\phi^a+\omega^a_{\;b}\phi^b] \\ \nonumber &=& d[d\phi^a+\omega^a_{\;b}\phi^b] +
\omega^a_{\;b}[d\phi^b+\omega^b_{\;c}\phi^c] \\ \nonumber &=& [d\omega^a_{\;b}
+\omega^a_{\;c}\wedge \omega^c_{\;b}]\phi^b.
\end{eqnarray}
The two-form within brackets in this last expression is a second rank Lorentz tensor (the curvature two-form)
\begin{eqnarray}\label{Curvature}
R_{\;b}^a &=& d\omega_{\;b}^a + \omega_{\;c}^a \wedge \omega_{\;b}^c. \\
\nonumber &=& \frac{1}{2}R_{\;b\mu \nu }^{a}dx^{\mu }\wedge dx^{\nu }
\end{eqnarray}

For a formal definition of this operator see, e.g., \cite{Eguchi-Gilkey-Hanson,Goeckeler-Schuecker}. The curvature two form defined by (\ref{Curvature}) is a Lorentz tensor on the tangent space, and is related to the Riemann tensor, $R_{\alpha\beta \mu \nu }$, through
\begin{equation}
R^{ab} = \frac{1}{2}e^a_{\;\alpha} e^b_{\;\beta}R^{\alpha\beta}_{\;\;\; \mu \nu}dx^{\mu} \wedge dx^{\nu}. \label{Riemann}
\end{equation}
This equivalence would not be true in a space with torsion, as we discuss in the next Appendix.

The fact that $\omega_{\;b}^{a}(x)$ and the gauge potential in Yang-Mills theory, $A_{\;b}^{a}=A_{\;b\mu }^{a}dx^{\mu }$, are both 1-forms and have similar properties is not an accident since they are both connections of a gauge group\footnote{In the precise language of mathematicians, $\omega$ is ``a locally defined Lie-algebra valued 1-form on $M$, which is also a connection in the principal $SO(D-1,1)$-bundle over $M$", while $A$ is ``a Lie-algebra valued 1-form on $M$, which is also a connection in the vector bundle of some gauge group $G$".}. Their transformation laws have the same form, and the curvature $R_{\;b}^{a}$ is completely analogous to the field strength in Yang-Mills, $\textbf{F}=d\textbf{A}+\textbf{A}\wedge \textbf{A}$.

\subsection*{A.4    Torsion}  
The fact that the two independent geometrical ingredients, $\omega$ and $e$, play different roles is underscored by their different transformation rules under the Lorentz group. In gauge theories this is reflected by the fact that vector fields play the role of matter, while the connection represents the carrier of interactions. 

Another important consequence of this asymmetry is the impossibility to construct a tensor two-form solely out of $e^a$ and its exterior derivatives, in contrast with the curvature which is uniquely defined by the connection. The only tensor obtained by differentiation of $e^{a}$ is its covariant derivative, also known the Torsion 2-form,
\begin{equation}
T^{a}=de^{a}+\omega^a_{\; b} \wedge e^{b},  \label{Torsion}
\end{equation}
which involves both the vielbein and the connection. In contrast with $T^a$, the curvature $R_{\;b}^{a}$ is not a covariant derivative of anything and depends only on $\omega$. In a manifold with torsion, one can split the connection into a torsion-free part and the so-called contorsion, $\omega=\bar{\omega} + \kappa$, where
\[ de^a + \bar{\omega}^a_{\, b}\wedge e^b \equiv 0, \quad \mbox{and} \quad  T^a = \kappa^a_{\, b}\wedge e^b.\]
In this case, the curvature two-form reads
\begin{equation}
R^a_{\; b}=\bar{R}^a_{\; b} + \bar{D}\kappa^a_{\; b},
\end{equation}
where $\bar{R}^a_{\, b}$ and $\bar{D}$ are the curvature and the covariant derivative constructed out of the torsion-free connection. It is the purely metric part of the curvature two-form, $\bar{R}^a_{\, b}$, that relates to the Riemann curvature through (\ref{Riemann}).

\subsection*{A.5    Bianchi identity} 

As we saw in \textbf{A.3}, taking the second covariant derivative of a vector amounts to multiplying by the curvature 2-form. As a consequence of a this relation between covariant differentiation and curvature, there exists an important property known as Bianchi identity,
\begin{equation}\label{Bianchi}
DR^{a}_{\;b} = dR_{\;b}^{a} + \omega_{\;c}^{a}\wedge R_{\;b}^{c} - \omega_{\;b}^{c}\wedge R_{\;c}^{a}\equiv 0\;.
\end{equation}
This is an identity and not a set of equations because it is satisfied for \textit{any} well defined connection 1-form whatsoever, and it does not restrict in any way the form of the field $\omega_{\;b \mu}^{a}$, which can be checked explicitly by substituting (\ref{Curvature}) in the second term of (\ref{Bianchi}). If conditions $DR_{\;b}^{a}=0$ were a set of equations instead, they would define a subset of connections that have a particular form, corresponding to some class of geometries.  

The Bianchi identity implies that the curvature $R^{ab}$ is ``transparent" for the exterior covariant derivative,
\begin{equation}
D(R^a_{\;b} \phi^b)= R^a_{\;b}\wedge D\phi^b, \;\;\; \mbox{and} \;\;\; DT^a=R^a_{\;b} \wedge e^b.
\end{equation} 

An important direct consequence of this identity is that by taking successive exterior derivatives of $e^a$, $\omega^{ab}$ and $T^a$ one does not generate new independent Lorentz tensors, in particular,
\begin{equation}
DT^a=R^a_{\;b} \wedge e^b.
\end{equation} 
The physical implication is that in the first order formulation, there is a very limited number of possible Lagrangians that can be constructed out of these fields in any given dimension \cite{Mardones-z}.


\end{document}